\documentclass[a4paper]{article}
\usepackage{geometry}
\usepackage{setspace}
\usepackage{float}
\usepackage{epstopdf}
\usepackage{standalone}
\usepackage{amsmath}

\usepackage{bm}
\usepackage{pdflscape}
\usepackage{authblk}
\usepackage{graphicx}
\usepackage[flushleft]{threeparttable}
\usepackage[comma,authoryear]{natbib}

\newenvironment{nalign}{
    \begin{equation}
    \begin{aligned}
}{
    \end{aligned}
    \end{equation}
    \ignorespacesafterend
}

\usepackage{mathtools}

\newcommand{\interior}[1]{%
 {\kern0pt#1}^{\mathrm{o}}%
}
\usepackage [english]{babel}
\usepackage [autostyle, english = american]{csquotes}
\MakeOuterQuote{"}

\usepackage{algorithm,algpseudocode}
\usepackage{caption}
\usepackage[makeroom]{cancel}

\newcommand{\E}{\mathrm{E}}
\newcommand{\Var}{\mathrm{Var}}
\newcommand{\Cov}{\mathrm{Cov}}

\title{\textbf{Reduced-dimensional Monte Carlo Maximum Likelihood for Latent Gaussian Random Field Models}}
\author[1]{Jaewoo Park}
\author[2]{Murali Haran}
\affil[1]{Department of Applied Statistics, Yonsei University}
\affil[2]{Department of Statistics, The Pennsylvania State University}

\begin{document}

\maketitle

\begin{abstract}

 Monte Carlo maximum likelihood (MCML) provides an elegant approach to find maximum likelihood estimators (MLEs) for latent variable models. However, MCML algorithms are computationally expensive when the latent variables are high-dimensional and correlated, as is the case for latent Gaussian random field models. Latent Gaussian random field models are widely used, for example in building flexible regression models and in the interpolation of spatially dependent data in many research areas such as analyzing count data in disease modeling and presence-absence satellite images of ice sheets. We propose a computationally efficient MCML algorithm by using a projection-based approach to reduce the dimensions of the random effects. We develop an iterative method for finding an effective importance function; this is generally a challenging problem and is crucial for the MCML algorithm to be computationally feasible. We find that our method is applicable to both continuous (latent Gaussian process) and discrete domain (latent Gaussian Markov random field) models. We illustrate the application of our methods to challenging simulated and real data examples for which maximum likelihood estimation would otherwise be very challenging. Furthermore, we study an often overlooked challenge in MCML approaches to latent variable models: practical issues in calculating standard errors of the resulting estimates, and assessing whether resulting confidence intervals provide nominal coverage. Our study therefore provides useful insights into the details of implementing MCML algorithms for high-dimensional latent variable models. 
\end{abstract}

\noindent%

\vfill

\clearpage
\section{Introduction}

~~~~~Monte Carlo maximum likelihood (MCML) provides an elegant and flexible approach for maximum likelihood estimation for latent variable models. However, MCML algorithms may not be practical in many settings,  especially in the context of high-dimensional latent variable models. For instance, they may be computationally infeasible for high-dimensional versions of stochastic volatility models \citep[cf.][]{jacquier2007mcmc,duan2017maximum} for financial econometrics, and finite Gaussian mixture models for galaxy data \citep{johansen2008particle}. A particularly important example of latent variable models where MCML estimation can be impractical is latent Gaussian random field models or spatial generalized linear mixed models (SGLMMs), which are popular models for spatial data, and widely used in many scientific domains \citep[cf.][]{banerjee2014hierarchical, lawson2016handbook}. The main challenge in MCML for high-dimensional latent variable models arises from the high-dimensional integration of a multivariate random variable. It is quite difficult to even obtain reasonable importance functions -- distributions from which samples are drawn to obtain a Monte Carlo approximate of the likelihood -- to carry out MCML for such models. In this manuscript, we develop a new computationally efficient approach for MCML for high-dimensional latent Gaussian random field models. Our method is based on projecting the high-dimensional latent variable to much lower dimensions, and involves a practical iterative approach for finding an importance function that leads to a computationally efficient algorithm. We also study different approaches for approximating the standard error of the maximum likelihood estimates. 

There is a vast literature on Bayesian approaches for latent Gaussian random field models, and there have been many algorithms proposed for handling the computational challenges effectively. Just a few of these include the predictive process \citep{banerjee2008gaussian}, the integrated nested Laplace approximation or INLA  \citep{rue2009approximate}, and reparameterization and dimension reduction methods \citep[cf.][]{haran2003accelerating,rue2005gaussian,christensen2006robust,hughes2013dimension,guan2018computationally}. Frequentist methods for latent Gaussian random field models have been much less popular than Bayesian approaches at least in part due to the fact that it is very computationally challenging to do maximum likelihood estimation in this context. Here, we focus on the challenge of making maximum likelihood estimation efficient for such models via the MCML approach. We note that a Bayesian approach typically allows more easily for flexible hierarchical modeling than frequentist methods, though  for certain commonly used SGLMMs a maximum likelihood approach can potentially lend itself to faster algorithms, and can also provide some automation by avoiding the need for prior specification. 

Several algorithms have been developed for maximum likelihood inference for SGLMMs. 
\cite{zhang2002estimation} proposes Monte Carlo expectation-maximization (MCEM) \citep{wei1990monte} and \cite{christensen2004monte} proposes Markov chain Monte Carlo maximum likelihood (MCML)  \citep{geyer1992constrained}. The MCEM algorithm uses Monte Carlo in the expectation step of the expectation-maximization (EM) algorithm, while the MCML algorithm works with a Monte Carlo approximation of the log likelihood function. These approximations require sampling the random effects and evaluating the likelihood function, which is computationally expensive for large data sets.  \cite{evangelou2011asymptotic, bonat2016practical} develop sophisticated Laplace approximations for the full conditional distribution of the random effects; this approach is related to INLA. However, both algorithms require finding optimized values of the random effects for Laplace approximations, which can be challenging for large data sets. Furthermore, unless the conditional distribution is well approximated by a Gaussian distribution, for high-dimensional data sets the Laplace approximation can be poor. Maximum likelihood inference has therefore rarely been used for analyzing large non-Gaussian spatial data sets. A recent exception is \cite{guan2019fast}  which develops a projection-based Monte Carlo Expectation-Maximization algorithm that is computationally feasible for such problems. However, as discussed in  \cite{knudson2016monte}, rigorously quantifying uncertainties for the estimates can be difficult, as MCEM algorithms lack the theoretical justifications of MCML. With increasing Monte Carlo sample size, the asymptotic properties of the likelihood approximations in MCML are well established \citep{geyer1992constrained,geyer1994convergence}. Specifically, with a compactness assumption for the parameter space, the MCMLE converges to the MLE almost surely \citep{geyer1994convergence}. The efficient MCML algorithm we develop in this manuscript can take advantage of existing asymptotic theories for estimating standard errors and establishing asymptotic confidence intervals. Notably, we show that a central limit theorem holds for the gradient of the Monte Carlo log likelihood approximation. This provides a rigorous justification for the convergence of MCMLE to the MLE; MCEM in general does not enjoy the theoretical convergence properties of MCMLE, though a notable exception is the theoretical work in \cite{fort2003convergence} on MCEM convergence for curved exponential families. As pointed out in \cite{knudson2016monte}, another advantage of the MCML compared with the MCEM is that MCML can provide log likelihood approximation without additional computational costs. This can be useful for likelihood-based inference beyond maximum likelihood. For example,  we can use the log likelihood approximations to calculate AIC for model selection. On the other hand, MCEM \citep{guan2019fast} can only evaluate the conditional expectations at specific parameter points without providing the log likelihood approximations. Furthermore, the MCML algorithm may be easier to extend to more general settings than the MCEM algorithm, for instance, when the distributions of the latent variables include intractable normalizing functions as in image denoising problems \citep{geman1986markov,tanaka2002statistical,murphy2012machine} or in hidden Markov models for disease mapping \citep{green2002hidden}.


In this manuscript we propose a fast MCML algorithm that uses projection methods \citep[cf.]{banerjee2012efficient,hughes2013dimension,guan2018computationally}. 
\cite{hughes2013dimension} develops the projection method for efficiently carrying out Bayesian inference. In this work, we develop a novel maximum likelihood approach based on MCML, which can fit latent Gaussian random field models quickly. Maximum likelihood inference for such models via standard MCML \citep{christensen2004monte} is computationally challenging since the evaluation of the likelihood function requires integration with respect to high-dimensional dependent latent variables. MCML methods, which enjoy nice theoretical properties \citep{geyer1992constrained,geyer1994convergence}, are not readily applicable to such problems. Our method is, to our knowledge, the first attempt to make MCML computationally feasible for fitting latent Gaussian random field models. While \cite{hughes2013dimension} focuses on the projection approach to areal data for the discrete spatial domain -- specifically the Gaussian Markov random field (GMRF) model -- the methods we develop here apply to a broader class of latent Gaussian fields, including both the GMRF and the Mat\'{e}rn class covariance function for point-referenced data. Therefore, our methods are broadly applicable to spatial data from both discrete and continuous domains, as we illustrate in several simulated and real data examples. We show that our algorithm can address the computational and inferential challenges of maximum likelihood estimation for SGLMMs. The outline of the remainder of this paper is as follows. In Section 2 we describe SGLMMs and introduce relevant notation, and in Section 3 we introduce a standard MCML approach and point out their computational challenges. In Section 4 we propose our fast dimension-reduced MCML approach to inference and provide implementation details. We also investigate in detail the computational complexity of our method. We study the performance of our approach via simulation studies in Section 5 and describe the application of our methods to two real data sets in Section 6. We conclude with a summary and discussion in Section 7.

\section{Spatial Generalized Linear Mixed Models}

~~~~~In this section, we describe spatial generalized linear mixed models (SGLMMs), which are latent Gaussian random field models for spatial data. Non-Gaussian spatial data are common in climate science, epidemiology, ecology and agriculture, among other disciplines. Examples include binary satellite data on ice sheets (ice-no ice) data \citep{chang2016calibrating}, and count data on plant disease \citep{zhang2002estimation}. Spatial generalized linear mixed models (SGLMMs) are popular and flexible models well suited for non-Gaussian spatial data, both for continuous domain \citep{diggle1998model} and discrete domain or lattice data \citep{besag1974spatial}. SGLMMs are commonly used to either interpolate these data or adjust for spatial dependence in a spatial regression model. However, for large data sets, maximum likelihood inference for such models becomes computationally very demanding because evaluating the likelihood function involves high-dimensional integration and operations on large matrices. Furthermore, spatial confounding between fixed and random effects can lead to unreliable parameter estimations \citep{reich2006effects,hanks2015restricted, guan2018computationally}. In this manuscript we provide a fast MCML algorithm for SGLMMs that also allows for adjustments to account for spatial confounding. 

We now describe SGLMMs, and provide relevant notation. Let $S \subset R^2$ be the spatial domain of interest. Consider $\mathbf{Z}= (Z(\mathbf{s}_{1}),\dots,Z(\mathbf{s}_{n}))$, the
observed data at coordinates $\mathbf{s} = (\mathbf{s}_{1},\dots,\mathbf{s}_{n}) \in S$ and  $\mathbf{X}=(x_{1}(\mathbf{s}),\dots,x_{p}(\mathbf{s})) \in R^{n \times p}$ is the matrix of covariates depending on the spatial locations. At each location we can define spatially correlated random effect $\mathbf{W}= (W(\mathbf{s}_{1}),\dots,W(\mathbf{s}_{n}))$. Then our model can be defined as follows. The random effects can be defined differently depending on whether the spatial domain of interest is continuous or discrete: (1) For continuous domains, $\mathbf{W}$ is a zero-mean second order stationary Gaussian process with a covariance kernel $\mathbf{C}_{\bm{\theta}}(W(\mathbf{s}+\mathbf{h}),W(\mathbf{s})) = \mathbf{C}_{\bm{\theta}}(\mathbf{h})$, where covariance function depends on the parameters $\bm{\theta}=(\sigma^2,\phi)$; $\sigma^2$ is the variance parameter and $\phi$ is the range parameter. For example, the Mat\'{e}rn class \citep{stein2012interpolation} covariance function is widely used. Then $\mathbf{W}$ follows a normal distribution, $f_{W}(\mathbf{W}|\bm{\theta}) \propto |\mathbf{C}_{\bm{\theta}}|^{-1/2}\exp(-\frac{1}{2}\mathbf{W}'\mathbf{C}_{\bm{\theta}}^{-1}\mathbf{W})$. (2) For discrete domains, $\mathbf{W}$ is assumed to follow a zero-mean Gaussian Markov random field (GMRF) \citep{besag1974spatial}. For all locations, if $i$th location and $j$th location are neighbors, $A_{i,j}=1$, otherwise $A_{i,j}=0$ and $A_{i,i}$ is defined as 0. This forms an $n \times n$ adjacency matrix $\mathbf{A}$. Then the distribution of $\mathbf{W}$ is $f_{W}(\mathbf{W}|\bm{\theta}) \propto \tau^{\mbox{rank}(\mathbf{Q})/2}\exp(-\frac{\bm{\theta}}{2}\mathbf{W'QW})$, where $\mathbf{Q}=$diag$(\mathbf{A}\bm{1})-\mathbf{A}$ and $\bm{\theta}=\tau$ controls the smoothness of the spatial field. 

Then with a link function $g(\cdot)$ the conditional mean $E[\mathbf{Z}|\mathbf{W},\bm{\beta}]$ can be modeled as 

\begin{equation}
g\lbrace E[\mathbf{Z}|\mathbf{W},\bm{\beta}]\rbrace  = \mathbf{X}\bm{\beta} + \mathbf{W}.
\label{linkfunction}
\end{equation}

\noindent Given random effects $\mathbf{W}$ and regression parameters $\bm{\beta}$, $\mathbf{Z}$ are assumed to be conditionally independent of each other (i.e. $f_{\mathbf{Z}|\mathbf{W}}(\mathbf{Z}|\mathbf{W},\bm{\beta}) = \prod_{i=1}^{n}f_{Z_{i}|W_{i}}(Z_{i}|W_{i},\bm{\beta})$). Let $\bm{\psi} = (\bm{\beta},\bm{\theta})$ be the model parameters for SGLMMs. Then the likelihood function is 

\begin{equation}
L(\bm{\psi}|\mathbf{Z}) = \int_{R^{n}}  f_{\mathbf{Z}|\mathbf{W}}(\mathbf{Z}|\mathbf{W},\bm{\beta})f_{\mathbf{W}}(\mathbf{W}|\bm{\theta})d\mathbf{W}.
\label{Likelihood}
\end{equation}

\noindent The number of random effects grows with the size of the data, which makes likelihood function evaluation difficult. Therefore, direct maximization of \eqref{Likelihood} is infeasible. 

We note that for SGLMMs, spatial confounding can lead to uninterpretability of parameters and variance inflation. Let $\mathbf{P}=\mathbf{X(X'X)^{-1}X'}$ and $\mathbf{P}^{\perp}= \mathbf{I}-\mathbf{P}$. Then our model can be represented as 
\begin{equation}
g\lbrace E[\mathbf{Z}|\mathbf{W},\bm{\beta}]\rbrace  = \mathbf{X}\bm{\beta} + \mathbf{W} = \mathbf{X}\bm{\beta} + \mathbf{P}\mathbf{W} + \mathbf{P}^{\perp}\mathbf{W}.
\label{confound}
\end{equation}

\noindent Then similar to the multicollinearity issue in standard regression models, $\mathbf{P}\mathbf{W}$ is confounded with $\mathbf{X}$ because they have a linear relationship. To alleviate spatial confounding, \cite{reich2006effects} suggests that $\mathbf{P}\mathbf{W}$ should be deleted from the model, creating what is called a restricted spatial regression model. For this model, \cite{hanks2015restricted} proposes to add a posteriori adjustment based on the MCMC samples of $\bm{\beta}$ to account effect of removing $\mathbf{P}\mathbf{W}$.  \cite{hughes2013dimension,guan2018computationally} develop fast reduced-dimensional Bayesian methods and \cite{guan2019fast} develops a Monte Carlo Expectation Maximization approach for maximum likelihood inference; all these approaches address both computational and confounding issues. In what follows, we develop a Monte Carlo maximum likelihood approach that is computationally efficient and also useful for addressing confounding issues.

\section{Monte Carlo Maximum Likelihood}

~~~~~Here we outline a Monte Carlo maximum likelihood (MCML) algorithm  \citep{geyer1992constrained, christensen2004monte} for SGLMMs. The likelihood function can be written as 

\begin{equation}
\begin{split}
L(\bm{\psi}|\mathbf{Z}) & =\int_{R^{n}}  f_{\mathbf{Z}|\mathbf{W}}(\mathbf{Z}|\mathbf{W},\bm{\beta})f_{\mathbf{W}}(\mathbf{W}|\bm{\theta})d\mathbf{W} = \int_{R^{n}} f_{\mathbf{Z},\mathbf{W}}(\mathbf{Z},\mathbf{W}|\bm{\psi})d\mathbf{W}\\
& \propto \int_{R^{n}} \frac{f_{\mathbf{Z},\mathbf{W}}(\mathbf{Z},\mathbf{W}|\bm{\psi})}{f_{\mathbf{Z},\mathbf{W}}(\mathbf{Z},\mathbf{W}|\widetilde{\bm{\psi}})}f_{\mathbf{W}|\mathbf{Z}}(\mathbf{W}|\mathbf{Z},\widetilde{\bm{\psi}})d\mathbf{W},
\label{MCLikelihood1}
\end{split}
\end{equation}

\noindent where $f_{\mathbf{W}|\mathbf{Z}}(\mathbf{W}|\mathbf{Z},\widetilde{\bm{\psi}}) \propto f_{\mathbf{Z}|\mathbf{W}}(\mathbf{Z}|\mathbf{W},\widetilde{\bm{\beta}})f_{\mathbf{W}}(\mathbf{W}|\widetilde{\bm{\theta})}$ is the conditional density for the  random effects with the model parameter $\widetilde{\bm{\psi}} = (\widetilde{\bm{\beta}}, \widetilde{\bm{\theta}})$. In the MCML context $f_{\mathbf{W}|\mathbf{Z}}(\mathbf{W}|\mathbf{Z},\widetilde{\bm{\psi}})$ is called the "importance function" and the importance function parameter $\widetilde{\bm{\psi}}$ should be reasonably close to the MLE for an accurate approximation. Then the MLE can be obtained by maximizing the following Monte Carlo approximation to the log likelihood function

\begin{equation}
\widehat{l}(\bm{\psi})=\log \left( \frac{1}{K} \sum_{k=1}^{K} \frac{f_{\mathbf{Z},\mathbf{W}}(\mathbf{Z},\mathbf{W}^{(k)}|\bm{\psi})}{f_{\mathbf{Z},\mathbf{W}}(\mathbf{Z},\mathbf{W}^{(k)}|\widetilde{\bm{\psi}})} \right),
\label{MCLikelihood2}
\end{equation}

\noindent where $\mathbf{W}^{(1)},\dots,\mathbf{W}^{(K)}$ are sampled by MCMC from the distribution $f_{\mathbf{W}|\mathbf{Z}}(\mathbf{W}|\mathbf{Z},\widetilde{\bm{\psi}})$. This approach works well for small data sets. However for more than thousands of data points, constructing efficient MCMC algorithms is challenging because the algorithms are expensive per update due to the dimensionality of the random effects. Furthermore, the Markov chains thus constructed tend to mix very slowly, which means that it takes many more iterations to obtain reasonable approximations. By using gradient information of the target posterior in the proposal distribution, a well mixing Langevin Hastings algorithm is proposed by \cite{christensen2002bayesian}, though the computational cost per iteration can increase significantly, thereby reducing the gains from better mixing. \cite{haran2003accelerating} and \cite{knorr2002block} suggest approaches for constructing proposals for large block updates of the random effects. Although this can improve the mixing of MCMC, application to large data sets can be challenging. Furthermore, the choice of importance function is crucial to the success of the MCML approach; that is, $\widetilde{\bm{\psi}}$ should be reasonably close to the MLE. To address these challenges, we propose a projection-based MCML approach in the following section. 

\section{Projection-based Monte Carlo Maximum Likelihood}

\subsection{Projection-based Dimension Reduction}

~~~~~Projection methods \citep{hughes2013dimension,guan2018computationally} have been developed for both continuous and discrete spatial domain to alleviate spatial confounding issues as well as to reduce computational costs. They reduce the dimension of the random effects from $n$ to $m(<<n)$ by using the $n \times m$ projection matrix $\mathbf{M}$.

In the continuous spatial domain, models include the zero-mean Gaussian random effects $\mathbf{W}$ with a covariance $\mathbf{C}_{\bm{\theta}} = \sigma^2\mathbf{R}_{{\phi}}$, where $\sigma^2$ is the variance parameter and $\phi$ is the range parameter. In the context of principal component analysis (PCA), the projection matrix $\mathbf{M}_{\phi}$ is constructed via $\mathbf{U_{\phi}D_{\phi}^{1/2}}$, where $\mathbf{U}_{\phi}$ is the first $m$ eigenvectors of correlation matrix $\mathbf{R}_{{\phi}}$, and $\mathbf{D}_{\phi}$ is a diagonal matrix with corresponding eigenvalues. The resulting model \citep{guan2018computationally} is

\begin{nalign} 
g\lbrace E[\mathbf{Z}|\bm{\beta},\mathbf{M}_{\phi},\bm{\delta}]\rbrace & = \mathbf{X}\bm{\beta} + \mathbf{M_{\phi}}\bm{\delta}\\
\bm{\delta} & \sim N(0,\sigma^2 \mathbf{I}).
\label{Reducsglmm}
\end{nalign}

\noindent By multiplying $\mathbf{P}^{\perp}$ to $\mathbf{M}_{\phi}$, a restricted model \citep{reich2006effects} can be easily constructed to eliminate spatial confounding. For every $\phi$ update, eigendecomposition on $\mathbf{R}_{{\phi}}$ is necessary. This can lead to computational challenges for large data sets.  \cite{guan2018computationally} replaces this exact eigendecomposition with a probabilistic version of  {N}ystr{\"o}m's approximation \citep{williams2001using,drineas2005nystrom}. We provide an outline of the random projection method in the supplementary material.

In the discrete case, we can construct $\mathbf{M}$ by taking the first $m$ principal components of the Moran operator $\mathbf{P}^{\perp}\mathbf{A}\mathbf{P}^{\perp}$, where $\mathbf{A}$ is an adjacency matrix. Such projection matrix can account for the spatial covariates of interest as well as for the underlying graph on a lattice domain. Then the projected model \citep{hughes2013dimension} can be written as follows.

\begin{nalign} 
g\lbrace E[\mathbf{Z}|\bm{\beta},\mathbf{M},\bm{\delta}]\rbrace & = \mathbf{X}\bm{\beta} + \mathbf{M}\bm{\delta}\\
f_{\bm{\delta}}(\bm{\delta}|\tau) & \propto \tau^{m/2}\exp(-\frac{\tau}{2}\bm{\delta}'\mathbf{M}'\mathbf{Q}\mathbf{M}\bm{\delta}).
\label{discreteReducsglmm}
\end{nalign}

In the following section we propose a projection-based MCML approach for SGLMMs. Our approach addresses computational and inferential challenges, and is generally applicable to both the continuous and discrete domain. Furthermore, we provide some guidance on how to tune the algorithm.

\subsection{Projection-based MCML}

~~~~~Here we propose a fast Monte Carlo maximum likelihood (MCML) algorithm for SGLMMs. Our approach can be applicable to both the  continuous and discrete spatial domain. Based on projection approaches in the previous section, the likelihood function can be represented as 

\begin{equation}
\begin{split}
L(\bm{\psi}|\mathbf{Z}) & = \int_{R^{m}} f_{\mathbf{Z}|\bm{\delta}}(\mathbf{Z}|\bm{\beta},\mathbf{M},\bm{\delta})f_{\bm{\delta}}(\bm{\delta}|\bm{\theta})d\bm{\delta} = \int_{R^{m}} f_{\mathbf{Z},\bm{\delta}}(\mathbf{Z},\bm{\delta}|\bm{\psi})d\bm{\delta}\\
& \propto \int_{R^{m}} \frac{f_{\mathbf{Z},\bm{\delta}}(\mathbf{Z},\bm{\delta}|\bm{\psi})}{f_{\mathbf{Z},\bm{\delta}}(\mathbf{Z},\bm{\delta}|\widetilde{\bm{\psi}})}f_{\bm{\delta}|\mathbf{Z}}(\bm{\delta}|\mathbf{Z},\widetilde{\bm{\psi}})d\bm{\delta},
\label{ReducLikelihood}
\end{split}
\end{equation}

\noindent where $f_{\bm{\delta}|\mathbf{Z}}(\bm{\delta}|\mathbf{Z},\widetilde{\bm{\psi}}) \propto f_{\mathbf{Z}|\bm{\delta}}(\mathbf{Z}|\widetilde{\bm{\beta}},\mathbf{M},\bm{\delta})f_{\bm{\delta}}(\bm{\delta}|\widetilde{\bm{\theta})}$ is the importance function for the reduced-dimensional  random effects. Then we can obtain the MLE by maximizing the following Monte Carlo approximation of the reduced-dimensional log likelihood function

\begin{equation}
\widehat{l}(\bm{\psi})=\log \left( \frac{1}{K} \sum_{k=1}^{K} \frac{f_{\mathbf{Z},\bm{\delta}}(\mathbf{Z},\bm{\delta}^{(k)}|\bm{\psi})}{f_{\mathbf{Z},\bm{\delta}}(\mathbf{Z},\bm{\delta}^{(k)}|\widetilde{\bm{\psi}})} \right),
\label{MCReducLikelihood}
\end{equation}

\noindent where $\bm{\delta}^{(1)},\dots,\bm{\delta}^{(K)}$ are sampled from the distribution $f_{\bm{\delta}|\mathbf{Z}}(\bm{\delta}|\mathbf{Z},\widetilde{\bm{\psi}})$ via MCMC. The MCML algorithm based on projection shows significant gains in computational efficiency over the original method  \citep{christensen2004monte}. This is because, compared to the full model likelihood function \eqref{MCLikelihood1}, \eqref{ReducLikelihood} requires a much smaller dimension integration ($m<<n)$. For example, in our simulation study we show that $m=50$ for $n=1,000$ can provide accurate estimates and prediction. Furthermore, $\bm{\delta}$ is less correlated than the original random effects $\mathbf{W}$, resulting in fast mixing of the MCMC for sampling  the random effects from $f_{\bm{\delta}|\mathbf{Z}}(\bm{\delta}|\mathbf{Z},\widetilde{\bm{\psi}})$. 

In the continuous spatial domain, model parameters are $\bm{\psi} = (\bm{\beta},\sigma^2,\phi)$. The maximization of \eqref{MCReducLikelihood} with respect to $\bm{\beta}$ and $\sigma^2$ is straightforward, because we can easily derive the first and second derivatives of \eqref{MCReducLikelihood}. Consider the conditional distribution of the response variable $f_{\mathbf{Z}|\bm{\delta}}(\mathbf{Z}|\bm{\beta},\mathbf{M},\bm{\delta})$ which is from the exponential family. Then the first and second derivatives with respect to the the regression parameters $\bm{\beta}$ are

\begin{nalign} 
\frac{\partial \log f_{\mathbf{Z}|\bm{\delta}}(\mathbf{Z}|\bm{\beta},\mathbf{M},\bm{\delta}) }{\partial \bm{\beta}} = \mathbf{X}'(\mathbf{Z}-E[\mathbf{Z}|\bm{\beta}, \mathbf{M}, \bm{\delta}])\\
\frac{\partial^2 \log f_{\mathbf{Z}|\bm{\delta}}(\mathbf{Z}|\bm{\beta},\mathbf{M},\bm{\delta}) }{\partial \bm{\beta}\partial \bm{\beta}'} =-\mathbf{X}'V(\mathbf{Z}|\bm{\beta}, \mathbf{M}, \bm{\delta})\mathbf{X},
\label{betaderiv}
\end{nalign}

\noindent where $V(\mathbf{Z}|\bm{\beta}, \mathbf{M}, \bm{\delta})$ is a diagonal matrix whose elements are the conditional variance of $\mathbf{Z}$. In the continuous case $\bm{\theta}=(\sigma^2,\phi)$ and derivatives with respect to the $\sigma^2$ are 

\begin{nalign} 
\frac{\partial \log f_{\bm{\delta}}(\bm{\delta}|\bm{\theta})}{\partial \sigma^2} =  -\frac{m}{2\sigma^2} + \frac{\bm{\delta}'\bm{\delta}}{2(\sigma^2)^2}\\
\frac{\partial^2 \log f_{\bm{\delta}}(\bm{\delta}|\bm{\theta})}{\partial (\sigma^2)^2} = \frac{m}{2(\sigma^2)^2} - \frac{\bm{\delta}'\bm{\delta}}{(\sigma^2)^3}.
\label{sigmaderivconti}
\end{nalign}

The derivatives with respect to $\phi$ do not have a closed form. For a given $\phi$, we use a Newton-Raphson procedure to obtain $\widehat{\bm{\beta}}(\phi)$ and $\widehat{\sigma}^2(\phi)$ which maximize \eqref{MCReducLikelihood}. We refer the reader to the supplementary material for calculating first and second derivatives of the objective function \eqref{MCReducLikelihood}. Then  $\widehat{\bm{\beta}}(\phi)$ and $\widehat{\sigma}^2(\phi)$ are plugged into \eqref{MCReducLikelihood} to obtain $\widehat{l}(\widehat{\bm{\beta}}(\phi),\widehat{\sigma}^2(\phi),\phi)$. This function is maximized with respect to $\phi$ using numerical optimization. Our projection-based MCML algorithm is described in Algorithm~\ref{MCMLContialg} below. 

\begin{algorithm}
\caption{Basic (non-iterative) projection-based MCML algorithm (continuous domain) }\label{MCMLContialg}
\begin{algorithmic}[H]
\normalsize
\State Given $\widetilde{\bm{\psi}} = ( \widetilde{\bm{\beta}}, \widetilde{\sigma}^2,\widetilde{\phi} )$ which is close to the MLE.

\State 1. Simulate random effects $\bm{\delta}^{(1)},\dots,\bm{\delta}^{(K)}$ via MCMC from $f_{\bm{\delta}|\mathbf{Z}}(\bm{\delta}|\mathbf{Z},\widetilde{\bm{\psi}})$.

\State 2. Construct a Monte Carlo approximation for the likelihood function as follows. 

$$\widehat{l}(\bm{\psi})=\log \left( \frac{1}{K} \sum_{k=1}^{K} \frac{f_{\mathbf{Z},\bm{\delta}}(\mathbf{Z},\bm{\delta}^{(k)}|\bm{\psi})}{f_{\mathbf{Z},\bm{\delta}}(\mathbf{Z},\bm{\delta}^{(k)}|\widetilde{\bm{\psi}})} \right)$$

\State 3. For given $\widetilde{\phi}$, obtain $\widehat{\bm{\beta}}(\widetilde{\phi}),\widehat{\sigma}^2(\widetilde{\phi})$ via an iterative Newton-Raphson method. 

\State 4. For given $\widehat{\bm{\beta}}(\widetilde{\phi})$ and $\widehat{\sigma}^2(\widetilde{\phi})$, obtain $\widehat{\phi}$ via a numerical optimization of $\widehat{l}(\widehat{\bm{\beta}}(\widetilde{\phi}),\widehat{\sigma}^2(\widetilde{\phi}),\phi)$.

\State 5. For given $\widehat{\phi}$, obtain $\widehat{\bm{\beta}}(\widehat{\phi}),\widehat{\sigma}^2(\widehat{\phi})$ via an iterative Newton-Raphson method. 

\end{algorithmic}
\end{algorithm}

In Step 1 of the Algorithm~\ref{MCMLContialg}, we construct a standard Metropolis-Hastings (MH) algorithm whose stationary distribution is $f_{\bm{\delta}|\mathbf{Z}}(\bm{\delta}|\mathbf{Z},\widetilde{\bm{\psi}})$ (see supplementary material for details). In practice, however, it is challenging to start the algorithm with $\widetilde{\bm{\psi}}$ which is reasonably close to the MLE. Therefore, we provide Algorithm~\ref{MCMLContialg2}, an iterative method for finding $\widetilde{\bm{\psi}}$. Starting with an arbitrary initial value $\bm{\psi}^{(0)}$ (e.g. GLM estimates), we repeat Algorithm~\ref{MCMLContialg} with one modification: we replace optimization of $\phi$ with numerical search on neighboring values of $\phi$ to reduce the computational expense of that step. An iterative search is repeated until $\widehat{l}_{t}(\bm{\psi}^{(t+1)})$ is close to 0, which implies that the $f_{\mathbf{Z},\bm{\delta}}(\mathbf{Z},\bm{\delta}^{(t,k)}|\bm{\psi}^{(t+1)})$ and $f_{\mathbf{Z},\bm{\delta}}(\mathbf{Z},\bm{\delta}^{(t,k)}|\bm{\psi}^{(t)})$ are close to each other. We present details about stopping rules for the algorithm in Section 4.3. 

\begin{algorithm}
\caption{An iterative method for finding $\widetilde{\bm{\psi}}$ (continuous domain)}\label{MCMLContialg2}
\begin{algorithmic}[H]
\normalsize
\State For $\bm{\psi}^{(t)} = (\bm{\beta}^{(t)}, \sigma^{2(t)},\phi^{(t)})$ at $t$th iteration following procedures are repeated until $\widehat{l}_{t}(\bm{\psi}^{(t+1)})$ becomes close to 0. 

\State 1. Simulate random effects $\bm{\delta}^{(t,1)},\dots,\bm{\delta}^{(t,K_{t})}$ via MCMC from $f_{\bm{\delta}|\mathbf{Z}}(\bm{\delta}|\mathbf{Z},\bm{\psi}^{(t)})$.

\State 2. Construct a Monte Carlo approximation for the objective function as follows. 

$$\widehat{l}_{t}(\bm{\psi})=\log \left( \frac{1}{K_{t}} \sum_{k=1}^{K_{t}} \frac{f_{\mathbf{Z},\bm{\delta}}(\mathbf{Z},\bm{\delta}^{(t,k)}|\bm{\psi})}{f_{\mathbf{Z},\bm{\delta}}(\mathbf{Z},\bm{\delta}^{(t,k)}|\bm{\psi}^{(t)})} \right)$$

\State 3. For $\phi^{(t)}$, obtain $\bm{\beta}(\phi^{(t)})^{(t+1)},\sigma^2(\phi^{(t)})^{(t+1)}$ via an iterative Newton-Raphson method. 

\State 4. For $\bm{\beta}(\phi^{(t)})^{(t+1)}$ and $\sigma^2(\phi^{(t)})^{(t+1)}$, obtain $\phi^{(t+1)}$ via a numerical search on the neighboring values of $\phi^{(t)}$ which maximizes $\widehat{l}_{t}(\bm{\beta}(\phi^{(t)})^{(t+1)},\sigma^2(\phi^{(t)})^{(t+1)},\phi)$.

\State 5. Set $(\bm{\beta}^{(t+1)}, \sigma^{2(t+1)},\phi^{(t+1)})=(\bm{\beta}(\phi^{(t)})^{(t+1)},\sigma^2(\phi^{(t)})^{(t+1)},\phi^{(t+1)})$

\end{algorithmic}
\end{algorithm}

In the discrete spatial domain, parameters of interest are $\bm{\psi}=(\bm{\beta},\tau)$. The derivatives with respect to $\bm{\beta}$ are identical to the derivatives in continuous case, \eqref{betaderiv}. However, derivatives with respect to the covariance parameter $\tau$ are derived separately because of difference in the covariance structure. The derivatives with respect to the $\tau$ are

\begin{nalign} 
\frac{\partial \log f_{\bm{\delta}}(\bm{\delta}|\tau)}{\partial \tau} =  \frac{m}{2\tau} - \frac{\bm{\delta}'\mathbf{M'QM}\bm{\delta}}{2}\\
\frac{\partial^2 \log f_{\bm{\delta}}(\bm{\delta}|\tau)}{\partial \tau^2} =  -\frac{m}{2\tau^2}.
\label{sigmaderivdis}
\end{nalign}

Since we have derivatives for both $\bm{\beta}$ and $\tau$, a projection-based MCML algorithm and its iterative search method may be simply written as shown in Algorithm~\ref{MCMLDisialg} and Algorithm~\ref{MCMLDisialg2}.

\begin{algorithm}
\caption{Basic (non-iterative) projection-based MCML algorithm (discrete domain) }\label{MCMLDisialg}
\begin{algorithmic}[H]
\normalsize
\State Given $\widetilde{\bm{\psi}} = ( \widetilde{\bm{\beta}}, \widetilde{\tau} )$ which is close to the MLE.

\State 1. Simulate random effects $\bm{\delta}^{(1)},\dots,\bm{\delta}^{(K)}$ via MCMC from  $f_{\bm{\delta}|\mathbf{Z}}(\bm{\delta}|\mathbf{Z},\widetilde{\bm{\psi}})$.

\State 2. Construct a Monte Carlo approximation for the likelihood function as follows. 

$$\widehat{l}(\bm{\psi})=\log \left( \frac{1}{K} \sum_{k=1}^{K} \frac{f_{\mathbf{Z},\bm{\delta}}(\mathbf{Z},\bm{\delta}^{(k)}|\bm{\psi})}{f_{\mathbf{Z},\bm{\delta}}(\mathbf{Z},\bm{\delta}^{(k)}|\widetilde{\bm{\psi}})} \right)$$

\State 3. Obtain $\widehat{\bm{\beta}},\widehat{\tau}$ via an iterative Newton-Raphson method. 

\end{algorithmic}
\end{algorithm}

\begin{algorithm}
\caption{An iterative method for finding $\widetilde{\bm{\psi}}$ (discrete domain)}\label{MCMLDisialg2}
\begin{algorithmic}[H]
\normalsize
\State For $\bm{\psi}^{(t)} = (\bm{\beta}^{(t)}, \tau^{(t)})$ at $t$th iteration following procedures are repeated until $\widehat{l}_{t}(\bm{\psi}^{(t+1)})$ becomes close to 0. 

\State 1. Simulate random effects $\bm{\delta}^{(t,1)},\dots,\bm{\delta}^{(t,K_{t})}$ via MCMC from $f_{\bm{\delta}|\mathbf{Z}}(\bm{\delta}|\mathbf{Z},\bm{\psi}^{(t)})$.

\State 2. Construct a Monte Carlo approximation for the objective function as follows. 

$$\widehat{l}_{t}(\bm{\psi})=\log \left( \frac{1}{K_{t}} \sum_{k=1}^{K_{t}} \frac{f_{\mathbf{Z},\bm{\delta}}(\mathbf{Z},\bm{\delta}^{(t,k)}|\bm{\psi})}{f_{\mathbf{Z},\bm{\delta}}(\mathbf{Z},\bm{\delta}^{(t,k)}|\bm{\psi}^{(t)})} \right)$$

\State 3. Obtain $\bm{\beta}^{(t+1)},\tau^{(t+1)}$ via an iterative Newton-Raphson method. 

\end{algorithmic}
\end{algorithm}

\subsection{Implementation Details}

~~~~~For our fast MCML approaches, we examine the approximation error. There are two sources of error for MCML estimates: (1) sampling error and (2) Monte Carlo error. \cite{knudson2016monte} derives two sources of error in the context of generalized linear mixed models (GLMMs). We apply the results in \cite{knudson2016monte} to our fast MCML approaches for SGLMMs. Consider an MLE $\widehat{\bm{\psi}}$ from the data with sample size $n$, and $\bm{\psi}^{\ast}$ is the true parameter value. Then under regularity conditions listed in  \cite{geyer2013asymptotics}, $\widehat{\bm{\psi}}$ asymptotically follows $N(\bm{\psi}^{\ast},[-\mathbf{A}]^{-1})$, where $\mathbf{A} = \nabla^2 l(\bm{\psi})$. The other source of error comes from the Monte Carlo approximation. Let $\widehat{\bm{\psi}}_{K}$ be the Monte Carlo estimates with $K$ number of Monte Carlo samples. Then under regularity conditions listed in \cite{geyer1994convergence} $\sqrt{K}(\widehat{\bm{\psi}}_{K}-\widehat{\bm{\psi}})$ converges to $N(0, \mathbf{A}^{-1}\mathbf{B}\mathbf{A}^{-1} )$, where $\mathbf{B}$ is
\begin{equation}
\mathbf{B} = \frac{   \int_{R^{m}} [\nabla \log f_{\mathbf{Z},\bm{\delta}}(\mathbf{Z},\bm{\delta}|\widehat{\bm{\psi}}][\nabla \log f_{\mathbf{Z},\bm{\delta}}(\mathbf{Z},\bm{\delta}|\widehat{\bm{\psi}}]'\frac{f_{\mathbf{Z},\bm{\delta}}(\mathbf{Z},\bm{\delta}|\widehat{\bm{\psi}})^2}{f_{\mathbf{Z},\bm{\delta}}(\mathbf{Z},\bm{\delta}|\widetilde{\bm{\psi}})^2}d\bm{\delta}  }{    \Big( \int_{R^{m}} f_{\mathbf{Z},\bm{\delta}}(\mathbf{Z},\bm{\delta}|\widehat{\bm{\psi}}) d\bm{\delta}  \Big)^2}
\label{MCerrorVariance}
\end{equation}
\noindent $\mathbf{A}$ can be estimated as $\nabla^2 \widehat{l}(\widehat{\bm{\psi}}_{K})$ based on the observed Fisher information. $\mathbf{B}$ can also be estimated as 
\begin{equation}
\widehat{\mathbf{B}} = \frac{\frac{1}{K}\sum_{k=1}^{K}[\nabla \log f_{\mathbf{Z},\bm{\delta}}(\mathbf{Z},\bm{\delta}^{(k)}|\widehat{\bm{\psi}}_{K})][\nabla \log f_{\mathbf{Z},\bm{\delta}}(\mathbf{Z},\bm{\delta}^{(k)}|\widehat{\bm{\psi}}_{K})]'\frac{f_{\mathbf{Z},\bm{\delta}}(\mathbf{Z},\bm{\delta}^{(k)}|\widehat{\bm{\psi}}_{K})^2}{f_{\mathbf{Z},\bm{\delta}}(\mathbf{Z},\bm{\delta}^{(k)}|\widetilde{\bm{\psi}})^2}}{\Big(\frac{1}{K}\sum_{k=1}^{K}\frac{f_{\mathbf{Z},\bm{\delta}}(\mathbf{Z},\bm{\delta}^{(k)}|\widehat{\bm{\psi}}_{K})}{f_{\mathbf{Z},\bm{\delta}}(\mathbf{Z},\bm{\delta}^{(k)}|\widetilde{\bm{\psi}})}\Big)^2}
\label{MCerrorVarianceEst}
\end{equation}

\noindent Then Monte Carlo errors can be obtained as $\widehat{\mathbf{A}}^{-1}\widehat{\mathbf{B}}\widehat{\mathbf{A}}^{-1}$. To justify the above results, we need to show a finite variance for $\nabla \widehat{l}(\psi)$. We provide the proof of this in the supplementary material by applying results in \cite{knudson2016monte}. The main idea of this is to bound a variance for $\nabla \widehat{l}(\psi)$ through the expectation of $f_{\mathbf{Z},\bm{\delta}}(\mathbf{Z},\bm{\delta}|\widetilde{\bm{\psi}})^{-1}$ with respect to a normal distribution. Once $\widetilde{\bm{\psi}}$ belongs to the compact region of the parameter space, $f_{\mathbf{Z},\bm{\delta}}(\mathbf{Z},\bm{\delta}|\widetilde{\bm{\psi}})^{-1}$ is bounded, because $f_{\mathbf{Z},\bm{\delta}}(\mathbf{Z},\bm{\delta}|\widetilde{\bm{\psi}})$ is the product of exponential family and normal densities. This is not a strong assumption, because we do not require $\widetilde{\bm{\psi}}$ to be close to the MLE in the proofs. In practice, however, $\widetilde{\bm{\psi}}$ should be reasonably close to the MLE for an accurate inference. Therefore, we use an iterative method as recommended in \cite{geyer1994convergence}. One might want to use this quantity as a stopping rule for the number of Monte Carlo samples. However, evaluating this value with each iteration of MCMC is computationally expensive because it requires that we calculate $\widehat{\bm{\psi}}_{K}$ for each $K$. Instead we use the Effective Sample Size (ESS) \citep{kass1998markov,robert2013monte} as a stopping rule. Incorporating these two sources of error is challenging in general. \cite{sung2007monte} found that the sampling error and Monte Carlo error can be added when the importance sampling distribution is constructed independently of the data. However we cannot apply this result directly because our importance sampling distribution $f_{\bm{\delta}|\mathbf{Z}}(\bm{\delta}|\mathbf{Z},\widetilde{\bm{\psi}})$ depends on the data. 

We can also obtain standard errors based on the parametric bootstrap \citep{efron1994introduction} as follows: (1) From the observed data, we obtain parameter estimates via our fast MCML approaches. (2) Multiple data sets are simulated from the SGLMMs for the estimated parameter values. (3) We obtain point estimates of simulated data sets using our fast MCML methods. (4) Finally we estimate standard errors from standard deviations of all parameter estimates. In this manuscript we compare the observed Fisher information-based standard errors with bootstrap-based standard errors. We found that bootstrap-based standard errors can provide more reasonable coverage (See Table~\ref{PoiSimulCoverage} and Table~\ref{BinarySimulCoverage}). 


For this projection-based MCML approach, we need to tune the following: (1) rank ($m$), (2) number of random effects simulations for Monte Carlo approximations, and (3) number of iterative search for finding $\widetilde{\bm{\psi}}$. For the rank selection we follow suggestions in \cite{hughes2013dimension,guan2018computationally}. The rank $m$ can be chosen from the desired proportion of variation (e.g. 95\%, explained by $\sum_{i=1}^{m}\lambda_{i}/\sum_{i=1}^{n}\lambda_{i}$). For the continuous domain we conduct eigendecomposition of the correlation matrix $\mathbf{R}_{\phi^{(0)}}$ for given initial $\phi^{(0)}$. Similarly, for the discrete domain we can take the first $m$ principal components of the Moran operator $\mathbf{P}^{\perp}\mathbf{A}^{\perp}\mathbf{P}^{\perp}$. Based on this, $m=50$ appears to suffice for all the examples we consider in our study.  As another option, \cite{guan2019fast} selects the rank based on a model selection criterion: we can fit non-spatial generalized linear models with predictors $\mathbf{X}$ and $\mathbf{M}_{\phi_0}$; here, $\mathbf{M}_{\phi_0}$ is the projection matrix for given initial $\phi^{(0)}$. We can fit several models with different $m$ values and compare AIC to determine the best model.

The number of MCMC iterations is determined by the effective sample size (ESS) \citep{kass1998markov,robert2013monte,gong2015practical}. ESS approximates the number of independent samples that corresponds to the number of dependent MCMC samples. For $K$ number of MCMC samples, ESS is  
\[
ESS(K)=\frac{K}{1+2\sum_{i=1}^{\infty}\rho_i},
\]
where $\rho_i$ is the autocorrelation of lag $i$. An almost independent sequence of samples would return an ESS similar to the actual length of the Markov chain; therefore, ESS provides a rough diagnostic for how well the chain is mixing. ESS has also been studied in the multivariate framework \citep{vats2019multivariate}. In Algorithm~1 and Algorithm~3 we run the MCMC until the ESS of the first random effect is at least 20 times the rank ($m$). In fact, our ESS-based stopping rule is equivalent to a relative standard deviation fixed width stopping rule \citep{gong2015practical}, which is theoretically justified. For $K$ number of Monte Carlo samples, \cite{gong2015practical} provides a stopping rule of $2z_{0.025}\widehat{\sigma}^2/\sqrt{K} \leq \widehat{\nu}\omega$, where $\widehat{\sigma}^2/\sqrt{K}$ is an estimated Monte Carlo standard errors and $\widehat{\nu}$ is an estimated posterior variance. This stopping criteria is equivalent to ESS $> 4z^2_{0.025}/\omega^2$. Based on this criteria for instance, $\omega=0.124$ can be converted to ESS $>1000$, which is identical to our stopping criteria of $20m$ for rank $m=50$. Considering that the projection methods reduce the dimension of random effects as well as make them uncorrelated, we believe that this criteria is conservative. 

We note that the Gelman-Rubin statistic \citep{gelman1992inference} is widely used to diagnose the convergence of MCMC. Recently, \cite{vats2018revisiting} improves the efficiency of the Gelman-Rubin statistic ($R$) based on the batch mean estimators \citep{vats2018lugsail}. Furthermore, they establish the one-to-one mapping between the Gelman-Rubin statistic ($R$) and ESS as
\[
R \approx \sqrt{1 + \frac{\mbox{number~of~parallel~chains}}{\mbox{ESS}}}.
\]
Based on this, for $m=50$ we use ESS$=20m=1000$ as a stopping criteria which yields Gelman-Rubin statistic with 10 parallel chains as 1.004988. This is much lower than the commonly used stopping criteria of 1.1. We note that the recommended cutoff of 1.1 is too high to provide reliable Monte Carlo estimates, as pointed out in \cite{vats2018revisiting}. However, for an iterative search (Algorithm~2 and Algorithm~4) we don't need a long run of MCMC, because Monte Carlo approximations only need to be accurate enough to obtain the improved importance function for each iteration. Therefore we run MCMC until the ESS of the first random effect is at least 3 times the rank. This yields Gelman-Rubin criteria as 1.032796, which is still a reasonably small threshold.  

There are two stopping rules for our algorithm. One is for deciding the number of MCMC samples (inner iteration) for log likelihood approximation, as we described above.  The other one is the number of iterative searches (outer iteration) for finding the importance function parameter $\widetilde{\psi}$.  In Algorithm~2 and Algorithm~4 iterative searches are repeated until $\widehat{l}_{t}(\bm{\psi}^{(t+1)})$ becomes close to 0, which indicates that the $f_{\mathbf{Z},\bm{\delta}}(\mathbf{Z},\bm{\delta}^{(t,k)}|\bm{\psi}^{(t+1)})$ and $f_{\mathbf{Z},\bm{\delta}}(\mathbf{Z},\bm{\delta}^{(t,k)}|\bm{\psi}^{(t)})$ are similar to each other. We stop the iterative search when $\widehat{l}_{t}(\bm{\psi}^{(t+1)}) + z_{0.05}ASE < \epsilon$, where ASE denotes the asymptotic Monte Carlo standard error of the sequence of log likelihood estimates $\lbrace \widehat{l}_{t}(\bm{\psi}^{(t+1)})\rbrace$ using batch means \citep{flegal2008markov}. We set $\epsilon$ to 0.5; this value was obtained by trial-and-error. We note that this is the stopping rule for the sequence of log likelihood approximation $\lbrace \widehat{l}_{t}(\bm{\psi}^{(t+1)}) \rbrace$, and is not for the MCMC samples. Based on this criteria we observe that $\widehat{l}_{t}(\bm{\psi}^{(t+1)})$ becomes close to 0 within several outer iterations. \cite{geyer1994convergence} also recommended such iterative searches for finding the improved importance function.

\subsection{Prediction}

~~~~~In many continuous spatial domain applications, prediction at unsampled locations is of great interest. Consider a prediction of locations $\mathbf{s}^{\ast} = (\mathbf{s}^{\ast}_{1},\dots,\mathbf{s}^{\ast}_{n^{\ast}}) \in S$. In the projection-based model the spatial random effects $\mathbf{W}$ at the observed coordinates can be reparameterized as $\mathbf{U_{\phi}D_{\phi}^{1/2}}\bm{\delta}$. To obtain $\mathbf{W}^{\ast}$ at some coordinates $\mathbf{s}^{\ast}$, we use basic definitions of the Gaussian process to obtain

\begin{equation}
\begin{bmatrix}
\mathbf{W} \\
\mathbf{W}^{\ast}
\end{bmatrix}
=MVN\left( \bm{0},
  \begin{bmatrix}
    \sigma^2 \mathbf{U}_{\phi}\mathbf{D}_{\phi}\mathbf{U}_{\phi}' & \mathbf{C}_{\bm{\theta},\mathbf{s}{\ast}}  \\
    \mathbf{C}_{\bm{\theta},{\ast}\mathbf{s}} & \mathbf{C}_{\bm{\theta},\ast\ast} 
  \end{bmatrix}
  \right),
\label{GPmat}
\end{equation}

\noindent where $\mathbf{C}_{\bm{\theta},\ast\mathbf{s}}=\mathbf{C}(\mathbf{s}^{\ast},\mathbf{s};\bm{\theta})$. Then the conditional distribution of $\mathbf{W}^{\ast}$ given  observed $\mathbf{W}$ is

\begin{equation}
\mathbf{W}^{\ast}|\mathbf{W} \sim MVN(  \frac{1}{\sigma^2}\mathbf{C}_{\bm{\theta},\ast\mathbf{s}}\mathbf{U}_{\phi}\mathbf{D}^{-1}_{\phi}\mathbf{U}_{\phi}'\mathbf{W},~~ \mathbf{C}_{\bm{\theta},\ast\ast}- \frac{1}{\sigma^2}\mathbf{C}_{\bm{\theta},\ast\mathbf{s}}\mathbf{U}_{\phi}\mathbf{D}^{-1}_{\phi}\mathbf{U}_{\phi}'\mathbf{C}_{\bm{\theta},\mathbf{s}{\ast}} ).
\label{condexp}
\end{equation}

To make a prediction of random effects, the Monte Carlo sample of the random effects $\mathbf{W}$ and the parameter estimates $\widehat{\bm{\theta}}$ from the last iteration of the MCML algorithm are plugged into \eqref{condexp}. Then we can obtain the empirical best linear unbiased predictor (EBLUP) of $\mathbf{W}^{\ast}$ from the mean in \eqref{condexp}.

\subsection{Computational Complexity}

~~~~~Here we examine the computational complexity of our projection-based MCML approaches and the standard MCML approaches \citep{christensen2004monte}, summarizing how our methods scale as we increase the size of the data. Table~\ref{Complexity} summarizes these calculations. We provide details about calculating computational complexity for the algorithms in the supplementary material.

\begin{table}[hh]
\centering
\begin{tabular}{cccc}
  \hline
Continuous Domain & Operations & Fast MCML & Standard MCML \\
  \hline
 & Importance sampling & $\mathcal{O}(nm)$   &   $\mathcal{O}(n^2)$ \\
 & Update $\bm{\beta},\sigma^2$ &  $\mathcal{O}(n^2)$ & $\mathcal{O}(n^2)$ \\
 & Update $\phi$ & $\mathcal{O}(nm^2)$ & $\mathcal{O}(n^3)$ \\
   \hline
Discrete Domain & Operations & Fast MCML & Standard MCML \\
  \hline
 & Importance sampling & $\mathcal{O}(nm)$  &   $\mathcal{O}(n^2)$ \\
 & Update $\bm{\beta},\tau$ &  $\mathcal{O}(n^2)$ & $\mathcal{O}(n^2)$\\
   \hline
\end{tabular}
\caption{Computational complexity. $n$ is the number of data points and $m$ is the reduced rank. }
\label{Complexity} 
\end{table}

The computational benefits come from reducing the dimension of random effects through projection methods. We could avoid expensive computations involving large covariance matrices when we update model parameters. Considering that $m$ is much smaller than $n$ (we use $m=50$ for thousands of $n$ in our study), our projection-based MCML approaches scales well compared to the standard MCML approaches. 

\section{Simulation Studies}

~~~~~We apply our approach to count and binary data for both the continuous and discrete domain. The code for this is implemented in {\tt R}. The source code can be downloaded from the following repository (https://github.com/jwpark88/projMCML). All the code was run on dual 10 core Xeon E5-2680 processors on the Penn State high performance computing cluster.

\subsection{Count Data in a Continuous Domain}

~~~~~We simulate a count data set with $n=1,400$ in the unit domain $[0,1]^2$ with model parameters $(\bm{\beta},\sigma^2,\phi)=(1,1,1,0.2)$. We first simulate random effects $\mathbf{W}$ using a Mat\'{e}rn class \citep{stein2012interpolation} covariance function with a smoothing parameter 2.5. Given simulated random effects, count observations are simulated from a Poisson distribution with rates $\bm{\lambda}=\exp{(\mathbf{X}\bm{\beta} + \mathbf{W})}$, where $\mathbf{X}$ is coordinate matrix for $\mathbf{W}$, randomly generated from the unit domain. We fit the model using the first 1,000 observations and 400 points are used for prediction. 

\begin{figure}
\begin{center}
\includegraphics[ scale = 0.6]{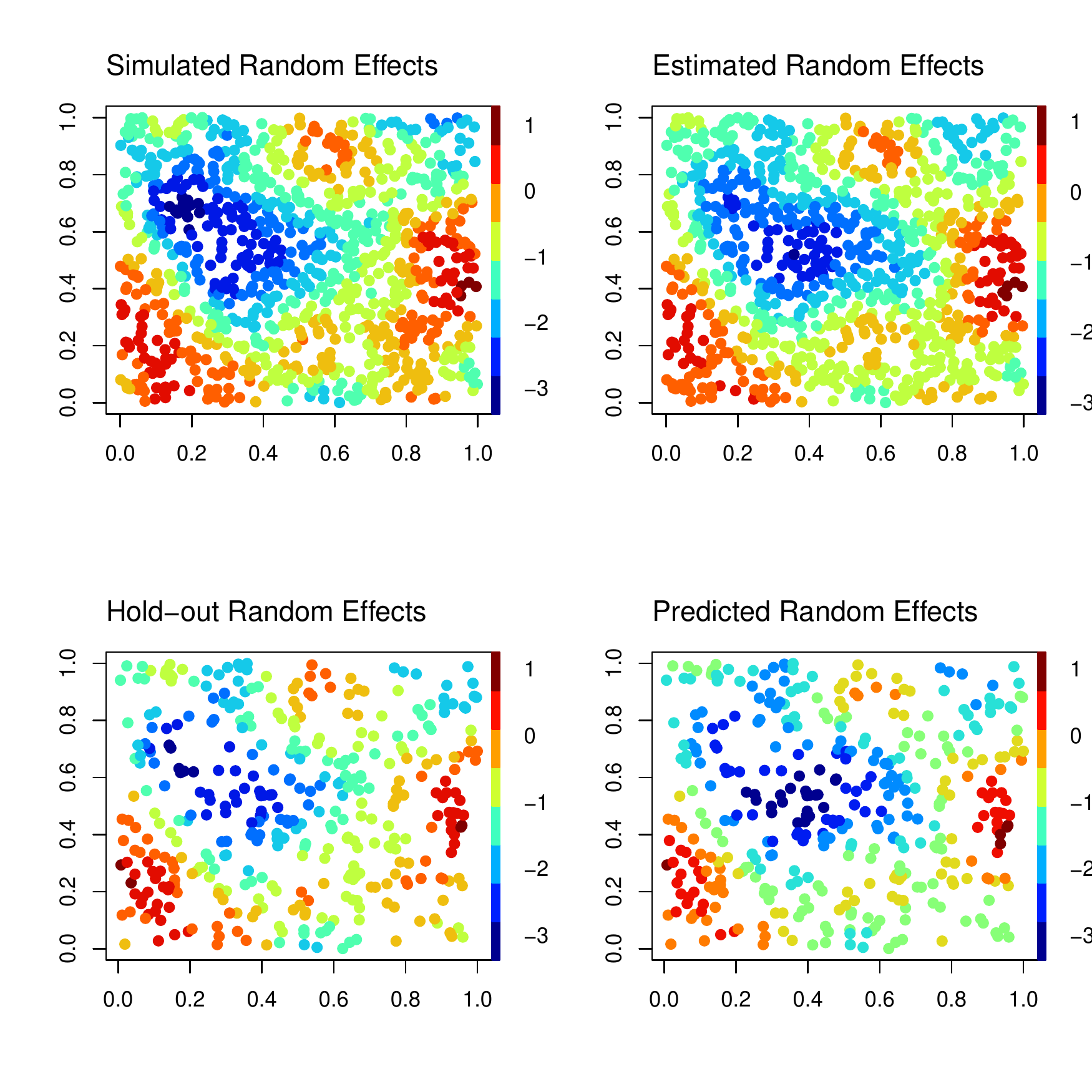}
\end{center}
\caption[]{The left panel shows the simulated random effects at the observation (top) and prediction locations (bottom). The right panel shows the estimated (top) and predicted (bottom) random effects. Spatial patterns are similar between simulated and predicted random effects. Estimations and predictions are obtained from the conditional distribution of the random effects using the point estimates from fast MCML (see Section 4.4 for details).}
\label{PoiSimulPred}
\end{figure}

Our initial values are obtained from GLM estimates. For the range parameter $\phi$, we use the first quartile of the Euclidean distance  between the locations. For the simulated example, this gives us initial values $\bm{\beta}^{(0)} = (0.546, 0.678)$ and $\bm{\theta}^{(0)}=(0.840, 0.333)$.
We select a reduced rank $m=50$ and follow the stopping criteria described in Section 4.3. Figure~\ref{PoiSimulPred} indicates agreement between simulated and predicted random effects; we observed similar spatial patterns. We study our methods for different values of $m$. Table~\ref{PoiSimulRank} shows that the parameter estimates from different choices of $m$ are reasonably close to the true values. We observed that the bootstrap-based confidence intervals are wider than the observed Fisher information-based confidence intervals. Furthermore, we can obtain confidence intervals for $\phi$ using bootstrap, without calculating analytically  intractable derivatives. 

\begin{table}
\centering
\begin{tabular}{ccccccc}
  \hline
m & 95\% CI & $\beta_{1}=1$ & $\beta_{2}=1$ & $\sigma^2=1$ & $\phi=0.2$ & Time(min)\\
  \hline
25 & & 0.996 &  0.922 & 0.999 & 0.190 & 7.676\\
  & Fisher  &  (0.859, 1.132) &  (0.784, 1.060) & (0.423, 1.576) &  NA\\
  & Bootstrap & (0.861, 1.127) & (0.745, 1.068) & (0.558, 5.734) &  (0.124, 0.411) \\
50 & & 1.078 & 0.952 & 1.53 & 0.237  & 17.450 \\ 
  & Fisher  &  (0.944, 1.212) &  (0.817, 1.086) & (0.877, 2.185) &  NA\\
  & Bootstrap & (0.922, 1.189) & (0.779, 1.054) & (0.659, 5.304) & (0.174, 0.384)  \\
75& & 1.030 & 0.912 &  1.48 & 0.255 & 23.439\\ 
  & Fisher  &  (0.971, 1.088) &  (0.877, 0.946) & (0.961, 1.992) &  NA\\
  & Bootstrap & (0.882, 1.144) & (0.767, 1.028) & (0.599,  3.799) & (0.168, 0.367)  \\
100 & & 1.030 & 0.901 &  1.44 & 0.246 & 31.774\\ 
  & Fisher  &  (0.872, 1.182) &  (0.746, 1.036) & (1.011, 1.872) &  NA\\
  & Bootstrap & (0.911, 1.169) & (0.726, 1.020) & (0.587, 3.846) & (0.179, 0.343)  \\
   \hline
\end{tabular}
\caption{Continuous domain simulated Poisson data; parameter estimates from different choices of $m$ are close to the true values. We cannot obtain standard errors of $\phi$ via Fisher information because it is analytically intractable.  }
\label{PoiSimulRank} 
\end{table}

\begin{figure}
\begin{center}
\includegraphics[ scale = 0.5]{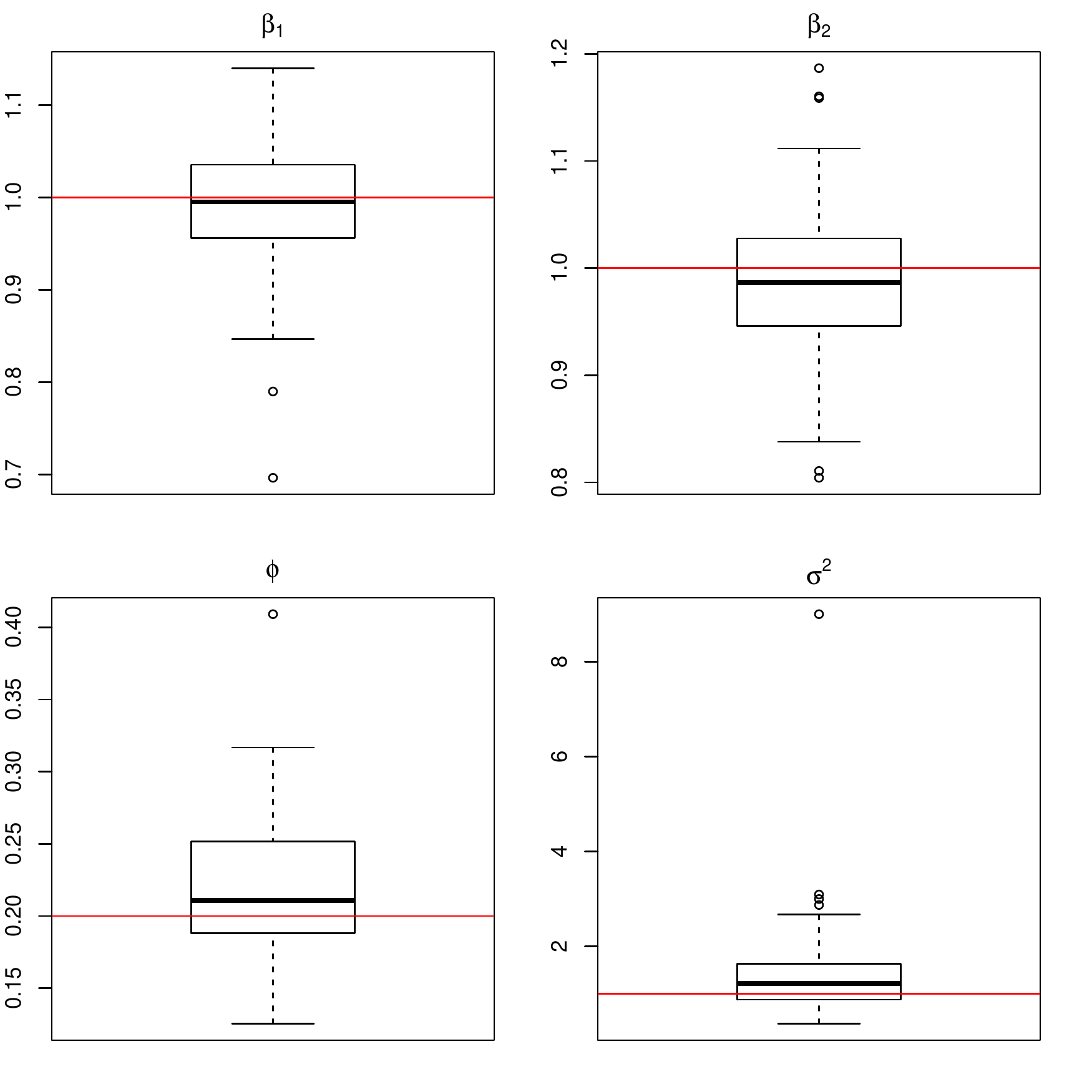}
\end{center}
\caption[]{Point estimates for the 100 simulated Poisson data sets in a continuous spatial domain. Horizontal lines indicate true value. Point estimates are distributed around the true values.}
\label{PoiSimul100}
\end{figure}

To validate our methods, we repeat the simulation 100 times. We used the same algorithm tuning parameters ($m=50$, $\epsilon=0.5$) for all the simulated data sets. Figure~\ref{PoiSimul100} indicates that the point estimates for each simulated data set are distributed around the true values. Compared to the regression parameters, estimates of covariance parameters are slightly right skewed, but still close to the true simulated values. We also investigate the coverages obtained from two different confidence intervals based on: (1) bootstrap, and (2) observed Fisher information. We observe that coverage based on the observed Fisher information is lower than the 95\% confidence. On the other hand,  coverage based on bootstrap are close to the nominal rate.

\begin{table}
\centering
\begin{tabular}{ccccc}
  \hline
Methods & $\beta_{1}$ & $\beta_{2}$ & $\sigma^2$ & $\phi$\\
  \hline
Fisher &  0.81 & 0.83 & 0.69 & NA\\
Bootstrap & 0.93 & 0.88 & 0.95 & 0.97 \\ 
   \hline
\end{tabular}
\caption{Coverage for simulated Poisson data sets in a continuous spatial domain.}
\label{PoiSimulCoverage} 
\end{table}

\subsection{Binary Data in a Continuous Domain}
 
~~~~~We simulate a binary data set with $n=1,400$ in the unit domain $[0,1]^2$ with parameters $(\bm{\beta},\sigma^2,\phi)=(1,1,1,0.2)$. Similar to the Poisson data simulation, we generate random effects $\mathbf{W}$ using a Mat\'{e}rn class covariance function with a smoothing parameter 2.5. Given simulated random effects, binary observations are simulated from a logit link function $\log \lbrace p/(1-p) \rbrace=\mathbf{X}\bm{\beta} + \mathbf{W}$, where $\mathbf{X}$ is coordinate matrix for $\mathbf{W}$. We use the first 1,000 observations and 400 points are used for prediction. 

\begin{figure}
\begin{center}
\includegraphics[ scale = 0.6]{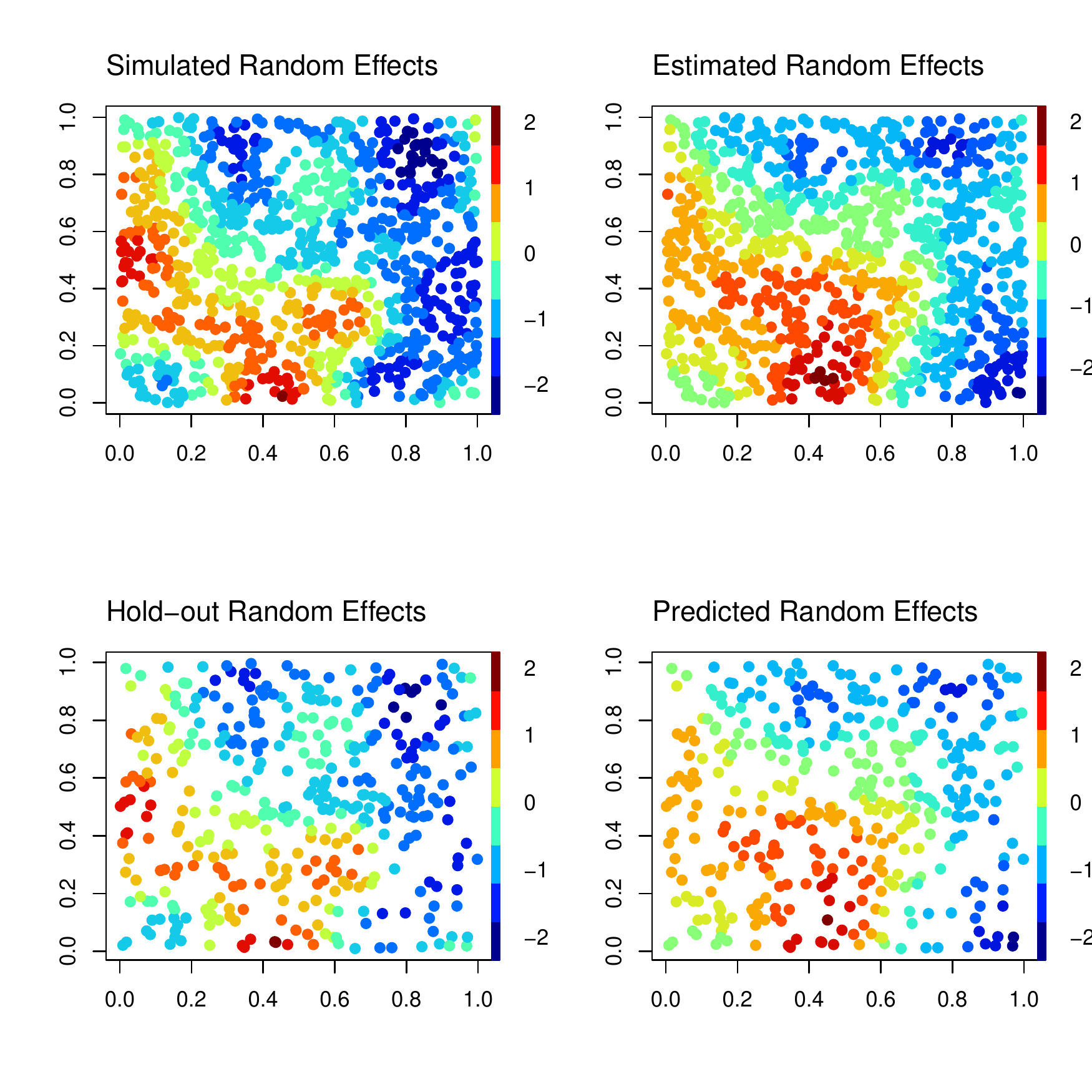}
\end{center}
\caption[]{The left panel shows the simulated random effects at the observation (top) and prediction locations (bottom). The right panel shows the estimated (top) and predicted (bottom) random effects. Spatial patterns are similar between simulated and predicted random effects. Estimations and predictions are obtained from the conditional distribution of the random effects using the point estimates from fast MCML (see Section 4.4 for details).}
\label{BinarySimulPred}
\end{figure}

We use initial values $\bm{\beta}^{(0)} = (0.352, 0.785)$ and $\bm{\theta}^{(0)}=(4.393, 0.327)$ which are obtained from GLM estimates. We choose a reduce rank $m=50$ and use stopping rules described in Section 4.3. We observed similar spatial patterns between simulated and predicted random effects (Figure~\ref{BinarySimulPred}). We also study our methods for different choices of $m$. In Table~\ref{BinarySimulRank}, we observed that Fisher information-based confidence intervals are wider than the those in the Poisson cases. Similar to the Poisson data examples, the bootstrap-based confidence intervals are wider than the observed Fisher information-based confidence intervals. We note that for this example we also have results from an MCEM algorithm \citep{guan2019fast}, which we provide for comparison: $(0.989,1.344,4.822,0.49)$ under the same simulation settings $(\bm{\beta},\sigma^2,\phi)=(1,1,1,0.2)$. 

\begin{table}
\centering
\begin{tabular}{ccccccc}
  \hline
m & 95\% CI & $\beta_{1}=1$ & $\beta_{2}=1$ & $\sigma^2=1$ & $\phi=0.2$ & Time(min)\\
  \hline
25 & & 0.869  & 1.405 & 2.274  & 0.292  & 6.919\\
  & Fisher & (0.520, 1.219)  & (0.994, 1.815) & (1.686, 2.862)  &  NA\\
  & Bootstrap & (-0.054, 1.403) & (0.807, 2.019) & (0.689, 32311657) & (0.117, 2.959) \\
50 & & 0.805 & 1.381 &  1.181  & 0.208 & 11.482 \\ 
  & Fisher & (0.403, 1.207) & (0.979, 1.783) & (0.646, 1.717) &  NA\\
  & Bootstrap & (0.227, 1.363)  & (0.851, 2.018) & (0.444, 13727.69) & (0.0989, 1.128)  \\
75& &  0.824 & 1.424 &  1.461 & 0.217 &  21.913\\ 
  & Fisher & (0.387, 1.259) & (0.995, 1.852) & (0.741, 2.181) &  NA\\
  & Bootstrap & (0.131, 1.505) & (0.832, 1.929) & (0.389, 192147.4) & (0.082, 1.101)  \\
100 & & 0.796 & 1.364 &  2.909 & 0.374 &  30.614\\ 
  & Fisher & (0.413, 1.180) & (0.973, 1.756) & (2.017, 3.802) &  NA\\
  & Bootstrap & (-0.003, 1.422) &  (0.714, 1.905) & (0.481, 559588) & (0.138, 7.918) \\
   \hline
\end{tabular}
\caption{Continuous domain simulated binary data; parameter estimates from different choices of $m$ are close to the true values. We cannot obtain standard errors of $\phi$ via Fisher information because it is analytically intractable.}
\label{BinarySimulRank} 
\end{table}

\begin{figure}
\begin{center}
\includegraphics[ scale = 0.5]{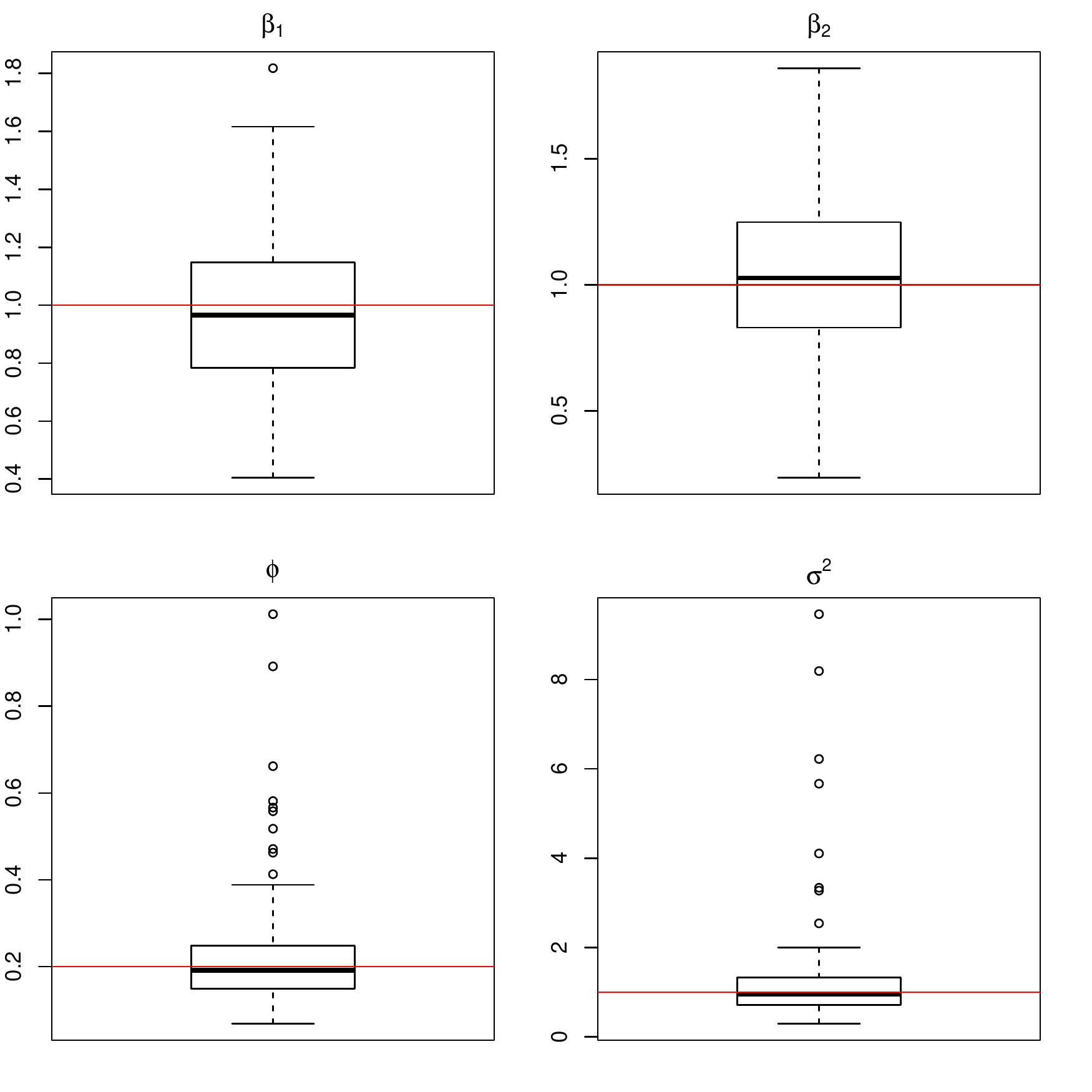}
\end{center}
\caption[]{Point estimates for the 100 simulated binary data sets in a continuous spatial domain. Horizontal lines indicate true value. Point estimates are distributed around the true values.}
\label{BinarySimul100}
\end{figure}

We repeat the simulation 100 times to investigate distribution of point estimates. It is observed that the point estimates are distributed around the true simulated values (Figure~\ref{BinarySimul100}). Coverage obtained from the observed Fisher information are higher than those in Poisson cases; but still lower than the nominal rate (95\% confidence). Coverage based on bootstrap are close to the nominal rate. 

\begin{table}
\centering
\begin{tabular}{ccccc}
  \hline
Methods & $\beta_{1}$ & $\beta_{2}$ & $\sigma^2$ & $\phi$\\
  \hline
Fisher  & 0.87 & 0.90 & 0.72 & NA\\
Bootstrap & 0.97 & 0.96 & 0.97 & 0.93\\ 
   \hline
\end{tabular}
\caption{Coverage for simulated binary data sets in a continuous spatial domain.}
\label{BinarySimulCoverage} 
\end{table}

\subsection{Count Data on a Lattice}

~~~~~Our methods are also applicable to the discrete spatial domain setting. We follow simulation settings in \cite{hughes2013dimension}. We create a Poisson data set on a $30 \times 30$ lattice over the unit domain with model parameters $(\bm{\beta},\tau)=(1,1,6)$. By taking first 400 eigenvectors of the Moran operator, we construct $900 \times 400$ matrix $\mathbf{M}$. Then we simulate random effects from $N(0,\tau\mathbf{M'QM})$. Given generated random effects, count observations are simulated from a Poisson distribution with rates $\bm{\lambda}=\exp{(\mathbf{X}\bm{\beta} + \mathbf{W})}$, where $\mathbf{X}$ is coordinate matrix for $\mathbf{W}$.

As in previous examples we use an initial value obtained from GLM estimates ($\bm{\beta}^{(0)} = (1.111, 0.989)$ and $\tau^{(0)}=2.498$). We choose a reduce rank $m=50$ and use stopping rules described in Section 4.3. We study our methods for different choices of $m$. In Table~\ref{PoiDiscreteSimulRank}, it is observed that all of the estimated values are close to the true simulated values. Both confidence intervals (Fisher information, bootstrap) are tighter than those in continuous domain cases. 

\begin{table}
\centering
\begin{tabular}{cccccc}
  \hline
m & 95\% CI & $\beta_{1}=1$ & $\beta_{2}=1$ & $\tau=6$ & Time(min)\\
  \hline
25 & & 1.069 & 0.963 & 6.263 &  9.558\\
  & Fisher  & (0.968, 1.171) & (0.861, 1.065) & (1.548, 10.979) \\
  & Bootstrap & (1.010, 1.153) & (0.906, 1.058) & (2.367, 9.775)  \\
50 & & 1.053 & 0.970 & 6.712 &  15.988\\ 
  & Fisher  & (0.951, 1.154) & (0.867, 1.072) & (2.680, 10.744)\\
  & Bootstrap & (0.986, 1.137) & (0.920, 1.049) & (3.117, 9.284) \\
75& &  1.049 & 0.964 & 6.94 & 23.684\\ 
  & Fisher  & (0.947, 1.152) & (0.861, 1.067) & (3.055, 10.819) \\
  & Bootstrap & (0.986, 1.144) & (0.913, 1.044) & (3.388, 8.435)  \\
100 & & 1.046 & 0.958 & 5.731 & 28.411\\ 
  & Fisher  & (0.944, 1.148) & (0.855, 1.061) & (2.780, 8.682) \\
  & Bootstrap & (0.992, 1.130) & (0.907, 1.032) & (2.916, 9.304) \\
   \hline
\end{tabular}
\caption{Discrete domain simulated Poisson data; parameter estimates from different choices of $m$ are close to the true values.}
\label{PoiDiscreteSimulRank} 
\end{table}

\begin{figure}
\begin{center}
\includegraphics[ scale = 0.8]{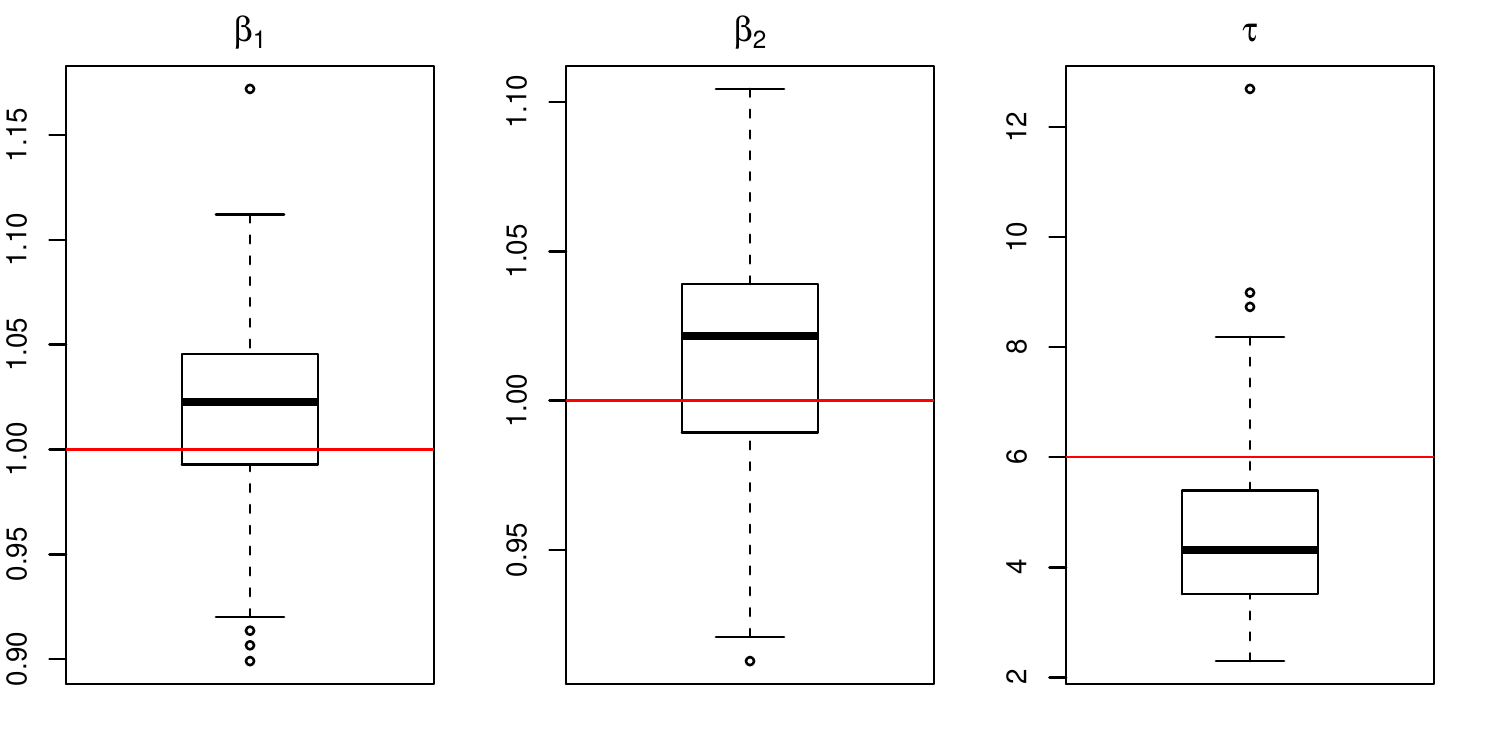}
\end{center}
\caption[]{Point estimates for the 100 simulated Poisson data sets in a lattice spatial domain. Horizontal lines indicate true value. Point estimates are distributed around the true values.}
\label{PoiDiscreteSimul100}
\end{figure}

We conduct the simulation 100 times, and the point estimates for $\bm{\beta}$ are distributed around the true simulated values. However the points estimates for $\tau$ are slightly biased (Figure~\ref{PoiDiscreteSimul100}). In Table~\ref{PoiDiscreteSimulCoverage}, it is observed that coverage obtained from the observed Fisher information and coverage based on the bootstrap are both close to the normal rate (95\% confidence). This is because we can simultaneously update $\bm{\beta}$ and $\tau$ based on derivatives with respect to each parameter unlike continuous domain cases. However coverage for $\tau$ are lower than nominal rate, due to the bias in point estimates. In general it is challenging to estimate $\tau$ correctly \citep{hughes2013dimension}.


\begin{table}
\centering
\begin{tabular}{cccc}
  \hline
Methods & $\beta_{1}$ & $\beta_{2}$ & $\tau$\\
  \hline
Fisher  & 0.93 & 0.96 & 0.63  \\
Bootstrap & 0.95 & 0.93 & 0.49  \\ 
   \hline
\end{tabular}
\caption{Coverage for simulated Poisson data sets in a discrete spatial domain.}
\label{PoiDiscreteSimulCoverage} 
\end{table}

\section{Real Data Examples}

~~~~~We illustrate the application of our approach to two real data examples: (1) binary mistletoe data over the continuous domain, and (2) county-level US infant mortality data. For both large non-Gaussian spatial data sets, a projection-based MCML algorithm can conduct statistical inference as well as provide prediction for unobserved locations within a reasonable amount of time. 

\subsection{Mistletoe Data}

~~~~~We apply our method to mistletoe data set from the Minnesota Department of Natural Resources forest inventory. The data set contains forest stand information from black spruce stands in Minnesota forests (the binary response indicates  whether or not dwarf mistletoe was found in the stand). Eastern spruce dwarf mistletoe (Arceuthobium pusillum) is a parasitic plant which causes the most serious disease of black spruce (Picea mariana) throughout its range \citep{baker2006eastern}. These data are analyzed in \cite{hanks2011reconciling} under a GLM framework. We use 4,000 locations as training and the use 1,000 locations for model prediction. In addition to coordinates (in Universal Transverse Mercator coordinate system), we also use average age of trees in stands, basal area per acre, average height of trees, and the total volume of trees in the stand as additional covariates. As in the simulated examples, we use initial values for model parameters from GLM estimates. We use an initial value of $\phi^{(0)}=39011.55$, which is the first quartile of the Euclidean distance matrix from coordinates. For a given initial value of $\phi^{(0)}$, we conduct eigendecomposition of the correlation matrix $\mathbf{R}_{\phi^{(0)}}$. We chose a reduced rank $m=50$, and this explains 99\% of variation ($\sum_{i=1}^{50}\lambda_{i}/\sum_{i=1}^{4000}\lambda_{i}=0.99$). We use stopping rules described in Section 4.3.

\begin{figure}
\begin{center}
\includegraphics[ scale = 0.77]{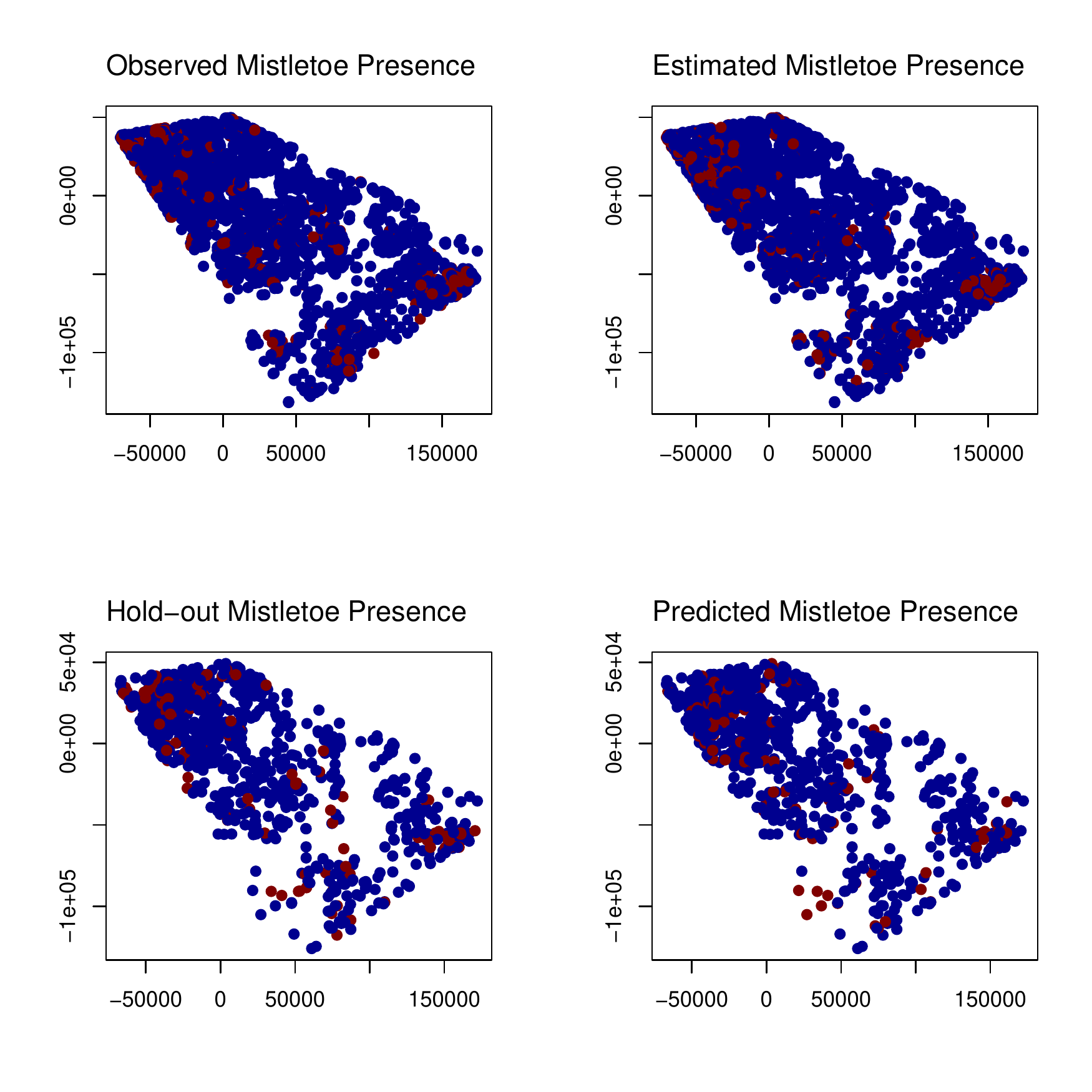}
\end{center}
\caption[]{The left panel shows the mistletoe presence at the observation (top) and prediction locations (bottom). The right panel shows the estimated (top) and predicted (bottom) mistletoe presence. Red and blue points indicate presence-absence of mistletoe respectively. Spatial patterns are similar between simulated and predicted random effects. Estimations and predictions are obtained from the conditional distribution of the random effects using the point estimates from fast MCML (see Section 4.4 for details).}
\label{mistletoePred}
\end{figure}

 Our fast MCML approach takes about 2 hours. Because of the expensive MCMC simulations to approximate the log likelihood function, regular MCML \citep{christensen2004monte} are impractical. 
  The projection-based MCEM approach in \citep{guan2019fast} may have comparable computational costs to our approach. However, MCML offers an alternative that has rigorous theoretical justifications  for the convergence of estimates as we described in Section 4.3. Also, as mentioned before, MCML provides Monte Carlo log likelihood approximations without additional effort, which is useful, for example, in popular model choice approaches like AIC. In contrast, MCEM \citep{guan2019fast} does not provide likelihood estimates as a by-product, because it only evaluates conditional expectations at several parameter points. Furthermore, our MCML approach may be more easily applicable to, and computationally tractable for, a broader class of models, for instance, image processing problems \citep{geman1986markov,tanaka2002statistical,murphy2012machine} or hidden Markov models for disease mapping \citep{green2002hidden}, where the distributions of the latent variables are intractable, for example when an Ising or Potts model is used as to model latent variables. Because our goal is maximum likelihood estimation, we do not compare our results to the plethora of fast algorithms for Bayesian inference, though we note that the Bayesian approaches based on similar projection-based methods are much more computationally expensive, for instance the algorithm in \citep{guan2018computationally} will take about several days to run.  
Inference results are summarized in Table~\ref{mistletoe}. Confidence intervals are obtained from the observed Fisher-information. 
Figure~\ref{mistletoePred} indicates that there are similar binary spatial patterns between observed and predicted locations.

\begin{table}[tt]
\centering
\begin{tabular}{ccc}
  \hline
Parameter & Estimate  & 95\% CI\\
  \hline
X-coordinates  & -2.242e-05 &  (-2.489e-05, -1.995e-05) \\
Y-coordinates  & 2.869e-06 & (-1.072e-06, 6.811e-06) \\
 Age  &  0.005 & (0.002, 0.007) \\
 Basal area  &  -0.002 & (-0.005, 0.001) \\
 Height  & 0.024 & (0.018, 0.030) \\
 Volume  &  -0.002 & (-0.004, 0.001) \\
 $\sigma^2$ & 4.772 & (2.893, 6.651) \\
 $\phi$ & 35847.78 & NA \\
   \hline
\end{tabular}
\caption{Inference results for a mistletoe data.}
\label{mistletoe} 
\end{table}

\subsection{US Infant Mortality Data}

~~~~~We apply our projection-based MCML algorithm to the infant mortality data set, which is observed in 3,071 US counties. The data set is obtained from the 2008 Area Resource File, which is maintained by the Bureau of Health Professions, Health Resources and Services Administration, US Department of Health and Human Services. This data set is analyzed in \cite{hughes2013dimension} under a Bayesian framework. As in \cite{hughes2013dimension}, we use the three-year (2002-2004) average number of infant deaths before the first birthday as a response variable. We use the rate of low birth weight, the percentage of black residents, the percentage of Hispanic, the Gini coefficient which measures income inequality, a composite score of social affluence, and residential stability as covariates (see \cite{hughes2013dimension} for details). Then, we use the average number of live births as an offset. As in the previous examples, we set the initial values for model parameters from GLM estimates. We chose a reduced rank $m=50$ as in \cite{hughes2013dimension}, and use same stopping rules as in Section 4.3. 

\begin{table}[tt]
\centering
\begin{tabular}{ccc}
  \hline
Parameter & Estimate  & 95\% CI\\
  \hline
Intercept & -5.423 & (-5.605, -5.240) \\
Low birth weight & 8.802 & (7.565, 10.038) \\
Black & 0.004 & (0.003, 0.006)\\
Hispanic & -0.004 & (-0.005, -0.003)\\
Gini & -0.575 & (-1.000, -0.151)\\
Affluence & -0.077 & (-0.089, -0.065) \\   
Stability & -0.029 & (-0.044, -0.015) \\ 
 $\tau$ & 7.815 &  (2.152, 13.479)\\
   \hline
\end{tabular}
\caption{Inference results for the US infant mortality data.}
\label{infant} 
\end{table}

Our approach takes about 1.5 hours. As in the binary mistletoe example, regular MCML \citep{christensen2004monte} is infeasible. In the continuous domain there is a signficant computational advantage of the MCML approach over the corresponding projection-based Bayesian approach via MCMC. In the discrete domain, however, the computational benefit of MCML are not as dramatic. The Bayes projection-based approach \citep{hughes2013dimension} takes about 3.5 hours for this example. However, a direct computational comparison is challenging because the Bayes approach is implemented in {\tt C++} using the Rcpp and RcppArmadillo packages \citep{eddelbuettel2011rcpp}. We note that \cite{hughes2013dimension} is applied to inference for latent GMRF models. However, our method is more general as it enables maximum likelihood inference for both the continuous and discrete spatial domains. Table~\ref{infant} summarizes inference results for the US infant mortality data set. We construct confidence intervals using the observed Fisher-information. The results in Table~\ref{infant} are comparable to those in \cite{hughes2013dimension}.

\section{Discussion}

~~~~~In this manuscript, we have proposed a fast Monte Carlo Maximum Likelihood approach for latent Gaussian random field models. Our methods are based on recently developed projection approaches \citep{hughes2013dimension,guan2018computationally}. By reducing the dimension of random effects, we can avoid expensive computations involving large covariance matrices as well as increase mixing of the algorithms for generating the MCMC samples for random effects, leading to a fast Monte Carlo approximation to the likelihood function. Our method is applicable to both the continuous and discrete domains. Our study shows that our approach obtains computational gains over existing methods and provides accurate point estimates. Our methods also provide accurate prediction in many applications (Figure~\ref{PoiSimulPred}, Figure~\ref{BinarySimulPred}, Figure~\ref{mistletoePred}). 

We have studied two sources of error for MCML estimates: sampling error and Monte Carlo error. Combining these two sources of error in cases where importance functions are dependent on the data is challenging \citep{knudson2016monte}. Asymptotic confidence intervals can be constructed using sampling error via the observed Fisher information. We observe that in the continuous domain cases, the coverage obtained from the observed Fisher information is lower than the nominal rate. These results are consistent with a well-known fact that, under fixed domain asymptotics, parameters in the Mat\'{e}rn class are not consistent and hence cannot be asymptotically normal \citep{zhang2004inconsistent}. As in the MCML algorithm for SGLMMs in  \cite{christensen2004monte}, our fast MCML approach cannot simultaneously update all model parameters $\psi=(\beta,\sigma^2,\phi)$ because the derivatives with respect to $\phi$ do not have a closed form. For a given $\phi$ our methods update $(\beta(\phi),\sigma^2(\phi))$ through the Newton-Raphson procedure using the observed Fisher information. Then, we update $\phi$ by maximizing approximated log likelihood using numerical optimization. These iterative procedures could result in lower coverage. Considering that interpolation is the ultimate goal in many geostatistical problems, constructing accurate confidence intervals is often of limited interest in practice. However, for researchers who have an interest in constructing confidence intervals, we recommend the bootstrap-based intervals, which have close to the nominal rate of coverage. We point out that the bootstrap-based approach is still computationally challenging, though we can take advantage of high-performance computing resources. To our knowledge, no existing approach provides practical ways to address this coverage issue in the context of SGLMMs. Developing practical ways for efficiently calculating confidence intervals for SGLMMs provides an interesting avenue for future research.

The computational methods developed in this manuscript allow researchers to perform maximum likelihood estimation to fit SGLMMs with much larger data sets than was previously possible, and we hope that they will permit researchers to fit such models more routinely. The algorithms we discuss here  apply to a large and widely applicable class of spatial models, and can be much faster than corresponding Bayesian approaches to fitting the same models. Bayesian approaches are of course enormously flexible, especially for handling missing data, incorporating prior information, and integrating multiple data sets;  there is an extensive literature on fast computational methods for fitting models in the Bayesian context \citep[cf.][]{lindgren2011explicit,banerjee2008gaussian,hughes2013dimension,guan2018computationally,rue2009approximate}.  The research on efficient MCMLE approaches for maximum likelihood estimation for computationally challenging problems is limited; we hope that the ideas and methods we have discussed here may be broadly relevant for efficient maximum likelihood inference for high-dimensional latent variable models. Examples include a Gaussian mixture model for galaxy data \citep{johansen2008particle}, stochastic volatility models in finance \citep{jacquier2007mcmc,duan2017maximum}, and state space models in economics \citep{duan2015density}.

\section*{Acknowledgement}
MH and JP were partially supported by the National Science Foundation through NSF-DMS-1418090. JP was partially supported by the Yonsei University Research Fund of 2019-22-0194 and the National Research Foundation of Korea (NRF-2020R1C1C1A01003868). Research reported in this publication was supported by the National Institute of General Medical Sciences of the National Institutes of Health under Award Number R01GM123007. The content is solely the responsibility of the authors and does not necessarily represent the official views of the National Institutes of Health. The authors are grateful to Yawen Guan and Christina Knudson for providing useful sample code and advice. 
\clearpage

\appendix
\begin{center}
\title{\LARGE\bf Supplementary Material for Reduced-dimensional Monte Carlo Maximum Likelihood for Spatial Generalized Linear Mixed Models}\\~\\
\author{\Large{Jaewoo Park and Murali Haran}}
\end{center}

\section{Monte Carlo Likelihood Approximation Calculations and Derivatives}

~~~~~Consider a logarithm of the Monte Carlo approximation to the likelihood as 

\begin{equation}
\widehat{l}(\bm{\psi}) := \log \widehat{L}(\bm{\beta},\bm{\theta};\mathbf{Z}) = \log \left( \frac{1}{K} \sum_{k=1}^{K} \frac{f_{\mathbf{Z},\bm{\delta}}(\mathbf{Z},\bm{\delta}^{(k)}|\bm{\psi})}{f_{\mathbf{Z},\bm{\delta}}(\mathbf{Z},\bm{\delta}^{(k)}|\widetilde{\bm{\psi}})} \right).
\label{MCMLapprox}
\end{equation}

\noindent The derivatives of the $f_{\mathbf{Z},\bm{\delta}}(\mathbf{Z},\bm{\delta}|\bm{\psi})$ can be defined as 

\begin{nalign} 
\nabla f_{\mathbf{Z},\bm{\delta}}(\mathbf{Z},\bm{\delta}|\bm{\psi}) =  \left(\frac{\partial \log f_{\mathbf{Z}|\bm{\delta}}(\mathbf{Z}|\bm{\beta},\mathbf{M},\bm{\delta}) }{\partial \bm{\beta}}, \frac{\partial \log f_{\bm{\delta}}(\bm{\delta}|\bm{\theta})}{\partial \bm{\theta}}\right)\\
\nabla^2 f_{\mathbf{Z},\bm{\delta}}(\mathbf{Z},\bm{\delta}|\bm{\psi}) =  \left(\frac{\partial^2 \log f_{\mathbf{Z}|\bm{\delta}}(\mathbf{Z}|\bm{\beta},\mathbf{M},\bm{\delta}) }{\partial \bm{\beta} \partial \bm{\beta}'}, \frac{\partial^2 \log f_{\bm{\delta}}(\bm{\delta}|\bm{\theta})}{\partial \bm{\theta}\partial \bm{\theta}'}\right).
\end{nalign}

\noindent From the chain rule, the gradient of the log likelihood can be derived as 

\begin{equation}
\nabla \widehat{l}(\bm{\psi}) = \frac{\frac{1}{K} \sum_{k=1}^{K} \frac{\nabla f_{\mathbf{Z},\bm{\delta}}(\mathbf{Z},\bm{\delta}^{(k)}|\bm{\psi})}{f_{\mathbf{Z},\bm{\delta}}(\mathbf{Z},\bm{\delta}^{(k)}|\widetilde{\bm{\psi}})}}{\frac{1}{K} \sum_{k=1}^{K} \frac{f_{\mathbf{Z},\bm{\delta}}(\mathbf{Z},\bm{\delta}^{(k)}|\bm{\psi})}{f_{\mathbf{Z},\bm{\delta}}(\mathbf{Z},\bm{\delta}^{(k)}|\widetilde{\bm{\psi}})}}  = \frac{ \sum_{k=1}^{K}[\nabla \log f_{\mathbf{Z},\bm{\delta}}(\mathbf{Z},\bm{\delta}^{(k)}|\bm{\psi}) ]\frac{f_{\mathbf{Z},\bm{\delta}}(\mathbf{Z},\bm{\delta}^{(k)}|\bm{\psi})}{f_{\mathbf{Z},\bm{\delta}}(\mathbf{Z},\bm{\delta}^{(k)}|\widetilde{\bm{\psi}})}}{ \sum_{k=1}^{K}\frac{f_{\mathbf{Z},\bm{\delta}}(\mathbf{Z},\bm{\delta}^{(k)}|\bm{\psi})}{f_{\mathbf{Z},\bm{\delta}}(\mathbf{Z},\bm{\delta}^{(k)}|\widetilde{\bm{\psi}})}}
\label{fristderiv}    
\end{equation}

\noindent By using the product rule, the Hessian of the MCML is as follows. 

\begin{equation}
\begin{split}
\nabla^2 \widehat{l}(\bm{\psi}) & =   
\frac{ \sum_{k=1}^{K}[\nabla^2 \log f_{\mathbf{Z},\bm{\delta}}(\mathbf{Z},\bm{\delta}^{(k)}|\bm{\psi}) ]\frac{ f_{\mathbf{Z},\bm{\delta}}(\mathbf{Z},\bm{\delta}^{(k)}|\bm{\psi}) }{ f_{\mathbf{Z},\bm{\delta}}(\mathbf{Z},\bm{\delta}^{(k)}|\widetilde{\bm{\psi}})} }{ \sum_{k=1}^{K}\frac{f_{\mathbf{Z},\bm{\delta}}(\mathbf{Z},\bm{\delta}^{(k)}|\bm{\psi})}{f_{\mathbf{Z},\bm{\delta}}(\mathbf{Z},\bm{\delta}^{(k)}|\widetilde{\bm{\psi}})} } + \frac{ \sum_{k=1}^{K}[\nabla \log f_{\mathbf{Z},\bm{\delta}}(\mathbf{Z},\bm{\delta}^{(k)}|\bm{\psi}) ]\frac{ \nabla f_{\mathbf{Z},\bm{\delta}}(\mathbf{Z},\bm{\delta}^{(k)}|\bm{\psi}) }{ f_{\mathbf{Z},\bm{\delta}}(\mathbf{Z},\bm{\delta}^{(k)}|\widetilde{\bm{\psi}})} }{ \sum_{k=1}^{K}\frac{f_{\mathbf{Z},\bm{\delta}}(\mathbf{Z},\bm{\delta}^{(k)}|\bm{\psi})}{f_{\mathbf{Z},\bm{\delta}}(\mathbf{Z},\bm{\delta}^{(k)}|\widetilde{\bm{\psi}})} } - \\
& \frac{ \sum_{k=1}^{K}[\nabla \log f_{\mathbf{Z},\bm{\delta}}(\mathbf{Z},\bm{\delta}^{(k)}|\bm{\psi}) ]\frac{f_{\mathbf{Z},\bm{\delta}}(\mathbf{Z},\bm{\delta}^{(k)}|\bm{\psi}) }{  f_{\mathbf{Z},\bm{\delta}}(\mathbf{Z},\bm{\delta}^{(k)}|\widetilde{\bm{\psi}})} }{  \Big( \sum_{k=1}^{K}\frac{f_{\mathbf{Z},\bm{\delta}}(\mathbf{Z},\bm{\delta}^{(k)}|\bm{\psi})}{f_{\mathbf{Z},\bm{\delta}}(\mathbf{Z},\bm{\delta}^{(k)}|\widetilde{\bm{\psi}})} \Big)^2 }\left( \nabla \sum_{k=1}^{K}\frac{f_{\mathbf{Z},\bm{\delta}}(\mathbf{Z},\bm{\delta}^{(k)}|\bm{\psi})}{f_{\mathbf{Z},\bm{\delta}}(\mathbf{Z},\bm{\delta}^{(k)}|\widetilde{\bm{\psi}})} \right) 
\label{secondderiv}
\end{split}
\end{equation}

\noindent Completing the derivative for each term gives
\begin{equation}
\begin{split}
\nabla^2 \widehat{l}(\bm{\psi}) & =   
\frac{ \sum_{k=1}^{K}[\nabla^2 \log f_{\mathbf{Z},\bm{\delta}}(\mathbf{Z},\bm{\delta}^{(k)}|\bm{\psi}) ]\frac{ f_{\mathbf{Z},\bm{\delta}}(\mathbf{Z},\bm{\delta}^{(k)}|\bm{\psi}) }{ f_{\mathbf{Z},\bm{\delta}}(\mathbf{Z},\bm{\delta}^{(k)}|\widetilde{\bm{\psi}})} }{ \sum_{k=1}^{K}\frac{f_{\mathbf{Z},\bm{\delta}}(\mathbf{Z},\bm{\delta}^{(k)}|\bm{\psi})}{f_{\mathbf{Z},\bm{\delta}}(\mathbf{Z},\bm{\delta}^{(k)}|\widetilde{\bm{\psi}})} } + \\
&  \frac{ \sum_{k=1}^{K}[\nabla \log f_{\mathbf{Z},\bm{\delta}}(\mathbf{Z},\bm{\delta}^{(k)}|\bm{\psi}) ][\nabla \log f_{\mathbf{Z},\bm{\delta}}(\mathbf{Z},\bm{\delta}^{(k)}|\bm{\psi}) ]'\frac{ f_{\mathbf{Z},\bm{\delta}}(\mathbf{Z},\bm{\delta}^{(k)}|\bm{\psi}) }{ f_{\mathbf{Z},\bm{\delta}}(\mathbf{Z},\bm{\delta}^{(k)}|\widetilde{\bm{\psi}})} }{ \sum_{k=1}^{K}\frac{f_{\mathbf{Z},\bm{\delta}}(\mathbf{Z},\bm{\delta}^{(k)}|\bm{\psi})}{f_{\mathbf{Z},\bm{\delta}}(\mathbf{Z},\bm{\delta}^{(k)}|\widetilde{\bm{\psi}})} } - \\
&  \frac{ \sum_{k=1}^{K}[\nabla \log f_{\mathbf{Z},\bm{\delta}}(\mathbf{Z},\bm{\delta}^{(k)}|\bm{\psi}) ]\frac{ f_{\mathbf{Z},\bm{\delta}}(\mathbf{Z},\bm{\delta}^{(k)}|\bm{\psi})}{f_{\mathbf{Z},\bm{\delta}}(\mathbf{Z},\bm{\delta}^{(k)}|\widetilde{\bm{\psi}})} }{ (\sum_{k=1}^{K}\frac{f_{\mathbf{Z},\bm{\delta}}(\mathbf{Z},\bm{\delta}^{(k)}|\bm{\psi})}{f_{\mathbf{Z},\bm{\delta}}(\mathbf{Z},\bm{\delta}^{(k)}|\widetilde{\bm{\psi}})}  )^2 }\left( \nabla \log f_{\mathbf{Z},\bm{\delta}}(\mathbf{Z},\bm{\delta}^{(k)}|\bm{\psi}) \right).
\label{secondderiv2}
\end{split}
\end{equation}

\noindent This can be simplified as 

\begin{equation}
\begin{split}
\nabla^2 \widehat{l}(\bm{\psi}) & =   
\frac{ \sum_{k=1}^{K}[\nabla^2 \log f_{\mathbf{Z},\bm{\delta}}(\mathbf{Z},\bm{\delta}^{(k)}|\bm{\psi}) ]\frac{ f_{\mathbf{Z},\bm{\delta}}(\mathbf{Z},\bm{\delta}^{(k)}|\bm{\psi}) }{ f_{\mathbf{Z},\bm{\delta}}(\mathbf{Z},\bm{\delta}^{(k)}|\widetilde{\bm{\psi}})} }{ \sum_{k=1}^{K}\frac{f_{\mathbf{Z},\bm{\delta}}(\mathbf{Z},\bm{\delta}^{(k)}|\bm{\psi})}{f_{\mathbf{Z},\bm{\delta}}(\mathbf{Z},\bm{\delta}^{(k)}|\widetilde{\bm{\psi}})} } +\\
& \frac{\sum_{k=1}^{K}[\nabla \log f_{\mathbf{Z},\bm{\delta}}(\mathbf{Z},\bm{\delta}^{(k)}|\bm{\psi}) - \nabla \widehat{l}(\bm{\psi}) ][\nabla \log f_{\mathbf{Z},\bm{\delta}}(\mathbf{Z},\bm{\delta}^{(k)}|\bm{\psi})- \nabla \widehat{l}(\bm{\psi})]'\frac{f_{\mathbf{Z},\bm{\delta}}(\mathbf{Z},\bm{\delta}^{(k)}|\bm{\psi})}{f_{\mathbf{Z},\bm{\delta}}(\mathbf{Z},\bm{\delta}^{(k)}|\widetilde{\bm{\psi}})} }{\sum_{k=1}^{K}\frac{f_{\mathbf{Z},\bm{\delta}}(\mathbf{Z},\bm{\delta}^{(k)}|\bm{\psi})}{f_{\mathbf{Z},\bm{\delta}}(\mathbf{Z},\bm{\delta}^{(k)}|\widetilde{\bm{\psi}})} }
\label{secondderiv3}
\end{split}
\end{equation}

\clearpage

\section{CLT for the gradient of the Monte Carlo log likelihood approximation}

Consider the gradient of the Monte Carlo log likelihood approximation 
\begin{equation}
\nabla \widehat{l}(\bm{\psi}) = \frac{ \frac{1}{K}\sum_{k=1}^{K}[\nabla \log f_{\mathbf{Z},\bm{\delta}}(\mathbf{Z},\bm{\delta}^{(k)}|\bm{\psi}) ]\frac{f_{\mathbf{Z},\bm{\delta}}(\mathbf{Z},\bm{\delta}^{(k)}|\bm{\psi})}{f_{\mathbf{Z},\bm{\delta}}(\mathbf{Z},\bm{\delta}^{(k)}|\widetilde{\bm{\psi}})}}{ \frac{1}{K}\sum_{k=1}^{K}\frac{f_{\mathbf{Z},\bm{\delta}}(\mathbf{Z},\bm{\delta}^{(k)}|\bm{\psi})}{f_{\mathbf{Z},\bm{\delta}}(\mathbf{Z},\bm{\delta}^{(k)}|\widetilde{\bm{\psi}})}},
\label{gradientMCL}    
\end{equation}
where $f_{\mathbf{Z},\bm{\delta}}(\mathbf{Z},\bm{\delta}|\bm{\psi}) \propto f_{\mathbf{Z}|\bm{\delta}}(\mathbf{Z}|\bm{\beta},\mathbf{M},\bm{\delta})f_{\bm{\delta}}(\bm{\delta}|\bm{\theta})$ is the joint distribution. Here, $f_{\mathbf{Z},\bm{\delta}}(\mathbf{Z},\bm{\delta}|\widetilde{\bm{\psi}})$ is bounded, if the $\widetilde{\bm{\psi}}$ belongs to the compact region of the parameter space. Then, the CLT for the gradient of the Monte Carlo log likelihood can be written as 
\begin{equation}
\sqrt{K}\Big[ \frac{ \frac{1}{K}\sum_{k=1}^{K}[\nabla \log f_{\mathbf{Z},\bm{\delta}}(\mathbf{Z},\bm{\delta}^{(k)}|\bm{\psi}) ]\frac{f_{\mathbf{Z},\bm{\delta}}(\mathbf{Z},\bm{\delta}^{(k)}|\bm{\psi})}{f_{\mathbf{Z},\bm{\delta}}(\mathbf{Z},\bm{\delta}^{(k)}|\widetilde{\bm{\psi}})}}{ \frac{1}{K}\sum_{k=1}^{K}\frac{f_{\mathbf{Z},\bm{\delta}}(\mathbf{Z},\bm{\delta}^{(k)}|\bm{\psi})}{f_{\mathbf{Z},\bm{\delta}}(\mathbf{Z},\bm{\delta}^{(k)}|\widetilde{\bm{\psi}})}} - \nabla l(\bm{\psi}) \Big]
\label{gradientCLT}, 
\end{equation}
which converges to a normal distribution with mean $0$ and variance
\begin{equation}
\frac{   \int_{R^{m}} [\nabla \log f_{\mathbf{Z},\bm{\delta}}(\mathbf{Z},\bm{\delta}|\widehat{\bm{\psi}}][\nabla \log f_{\mathbf{Z},\bm{\delta}}(\mathbf{Z},\bm{\delta}|\widehat{\bm{\psi}}]'\frac{f_{\mathbf{Z},\bm{\delta}}(\mathbf{Z},\bm{\delta}|\widehat{\bm{\psi}})^2}{f_{\mathbf{Z},\bm{\delta}}(\mathbf{Z},\bm{\delta}|\widetilde{\bm{\psi}})^2}d\bm{\delta}  }{    \Big( \int_{R^{m}} f_{\mathbf{Z},\bm{\delta}}(\mathbf{Z},\bm{\delta}|\widehat{\bm{\psi}}) d\bm{\delta}  \Big)^2}.
\label{gradientVar}
\end{equation}
To verify this, we need to show \eqref{gradientVar} is finite. 
This requires to show both the numerator of \eqref{gradientMCL}
\[
\frac{1}{K}\sum_{k=1}^{K}[\nabla \log f_{\mathbf{Z},\bm{\delta}}(\mathbf{Z},\bm{\delta}^{(k)}|\bm{\psi}) ]\frac{f_{\mathbf{Z},\bm{\delta}}(\mathbf{Z},\bm{\delta}^{(k)}|\bm{\psi})}{f_{\mathbf{Z},\bm{\delta}}(\mathbf{Z},\bm{\delta}^{(k)}|\widetilde{\bm{\psi}})}
\]
and denominator of \eqref{gradientMCL}
\[
\frac{1}{K}\sum_{k=1}^{K}\frac{f_{\mathbf{Z},\bm{\delta}}(\mathbf{Z},\bm{\delta}^{(k)}|\bm{\psi})}{f_{\mathbf{Z},\bm{\delta}}(\mathbf{Z},\bm{\delta}^{(k)}|\widetilde{\bm{\psi}})},
\]
have a finite variance respectively. In what follows we apply the results in \cite{knudson2016monte} to our fast MCML methods for SGLMMs. 

\clearpage
\subsection{A Finite Variance for the Numerator Part}

We need to show the following three conditions
\begin{equation}
\Var\Big( \frac{\partial \log f_{\mathbf{Z}|\bm{\delta}}(\mathbf{Z}|\bm{\beta},\mathbf{M},\bm{\delta}) }{\partial \bm{\beta}}   \frac{f_{\mathbf{Z},\bm{\delta}}(\mathbf{Z},\bm{\delta}|\bm{\psi})}{f_{\mathbf{Z},\bm{\delta}}(\mathbf{Z},\bm{\delta}|\widetilde{\bm{\psi}})} \Big) < \infty    
\label{cond1}
\end{equation}
\begin{equation}
\Var\Big(\frac{\partial \log f_{\bm{\delta}}(\bm{\delta}|\bm{\theta})}{\partial \bm{\theta}}  \frac{f_{\mathbf{Z},\bm{\delta}}(\mathbf{Z},\bm{\delta}|\bm{\psi})}{f_{\mathbf{Z},\bm{\delta}}(\mathbf{Z},\bm{\delta}|\widetilde{\bm{\psi}})} \Big) < \infty    
\label{cond2}
\end{equation}
\begin{equation}
\Big|\Cov\Big(\frac{\partial \log f_{\mathbf{Z}|\bm{\delta}}(\mathbf{Z}|\bm{\beta},\mathbf{M},\bm{\delta}) }{\partial \bm{\beta}}   \frac{f_{\mathbf{Z},\bm{\delta}}(\mathbf{Z},\bm{\delta}|\bm{\psi})}{f_{\mathbf{Z},\bm{\delta}}(\mathbf{Z},\bm{\delta}|\widetilde{\bm{\psi}})}, 
\frac{\partial \log f_{\bm{\delta}}(\bm{\delta}|\bm{\theta})}{\partial \bm{\theta}}  \frac{f_{\mathbf{Z},\bm{\delta}}(\mathbf{Z},\bm{\delta}|\bm{\psi})}{f_{\mathbf{Z},\bm{\delta}}(\mathbf{Z},\bm{\delta}|\widetilde{\bm{\psi}})}
\Big)\Big| < \infty    
\label{cond3}
\end{equation}
Since \eqref{cond1} and \eqref{cond2} imply \eqref{cond3} by the Cauchy-Schwartz inequality, we only need to show \eqref{cond1} and \eqref{cond2}. For each $\beta_j$, 
\begin{equation}
\begin{split}
\Var\Big( \frac{\partial \log f_{\mathbf{Z}|\bm{\delta}}(\mathbf{Z}|\beta_j,\mathbf{M},\bm{\delta}) }{\partial \beta_j}   \frac{f_{\mathbf{Z},\bm{\delta}}(\mathbf{Z},\bm{\delta}|\bm{\psi})}{f_{\mathbf{Z},\bm{\delta}}(\mathbf{Z},\bm{\delta}|\widetilde{\bm{\psi}})} \Big) & <
\E\Big[\Big( \frac{\partial \log f_{\mathbf{Z}|\bm{\delta}}(\mathbf{Z}|\beta_j,\mathbf{M},\bm{\delta}) }{\partial \beta_j}   \frac{f_{\mathbf{Z},\bm{\delta}}(\mathbf{Z},\bm{\delta}|\bm{\psi})}{f_{\mathbf{Z},\bm{\delta}}(\mathbf{Z},\bm{\delta}|\widetilde{\bm{\psi}})} \Big)^2\Big] \\
& = \int \Big[ \frac{\partial \log f_{\mathbf{Z}|\bm{\delta}}(\mathbf{Z}|\beta_j,\mathbf{M},\bm{\delta}) }{\partial \beta_j}  \Big]^2
\frac{[f_{\mathbf{Z}|\bm{\delta}}(\mathbf{Z}|\beta_j,\mathbf{M},\bm{\delta})]^2[f_{\bm{\delta}}(\bm{\delta}|\bm{\theta})]^2}{f_{\mathbf{Z},\bm{\delta}}(\mathbf{Z},\bm{\delta}|\widetilde{\bm{\psi}})} 
d\bm{\delta}.
\label{cond1bound1}
\end{split}
\end{equation}
Since $f_{\mathbf{Z}|\bm{\delta}}(\mathbf{Z}|\beta_j,\mathbf{M},\bm{\delta})$ belongs to the exponential family, it can be written as $\exp(\mathbf{Z}'\bm{\eta} - b(\bm{\eta}))$ for $\bm{\eta}=\mathbf{X}\bm{\beta} + \mathbf{M}\bm{\delta}$. Let $X_{\cdot j}$ be the $j$th column of $\mathbf{X}$. Then, we have 
\[
\frac{\partial \log f_{\mathbf{Z}|\bm{\delta}}(\mathbf{Z}|\beta_j,\mathbf{M},\bm{\delta}) }{\partial \beta_j} = X_{\cdot, j}'(\mathbf{Z}-b'(\bm{\eta})).
\]
Also, $f_{\bm{\delta}}(\bm{\delta}|\bm{\theta}) = c\exp(-\bm{\delta}'\bm{\delta}/2\bm{\theta})$ for some constant $c$. By plugging in these into \eqref{cond1bound1}, we have
\begin{equation}
c\int [X_{\cdot, j}'(\mathbf{Z}-b'(\bm{\eta}))]^2 
\frac{\exp(2\mathbf{Z}'\eta)f_{\mathbf{Z},\bm{\delta}}(\mathbf{Z},\bm{\delta}|\widetilde{\bm{\psi}})^{-1}}{\exp(2b(\bm{\eta}))\exp(\bm{\delta}'\bm{\delta}/\bm{\theta})}
d\bm{\delta}
\label{cond1bound2}
\end{equation}
Since $0 \leq \exp(-2b(\bm{\eta}) )\leq 1$ for both Bernoulli and Poisson cases, \eqref{cond1bound2} is bounded by  
\begin{equation}
c\exp(2\mathbf{Z}'\mathbf{X}\bm{\beta})\int [X_{\cdot, j}'(\mathbf{Z}-b'(\bm{\eta}))]^2 
\frac{\exp(2\mathbf{Z}'\mathbf{M}\delta)f_{\mathbf{Z},\bm{\delta}}(\mathbf{Z},\bm{\delta}|\widetilde{\bm{\psi}})^{-1}}{\exp(\bm{\delta}'\bm{\delta}/\bm{\theta})}
d\bm{\delta}
\label{cond1bound3}
\end{equation}
For the Bernoulli response case, by the triangle inequality and with some algebra, \eqref{cond1bound3} is bounded by
\begin{equation}
c\exp(2\mathbf{Z}'\mathbf{X}\bm{\beta})\Big[\sum_{i=1}^{n}|x_{ij}| \Big]^2\int f_{\mathbf{Z},\bm{\delta}}(\mathbf{Z},\bm{\delta}|\widetilde{\bm{\psi}})^{-1}\exp(-\bm{\delta}'\bm{\delta}/\bm{\theta} + 2\mathbf{Z}'\mathbf{M}\delta)d\bm{\delta}
\label{cond1bound4}
\end{equation}
Here, \eqref{cond1bound4} is proportional to the expectation of the bounded function $f_{\mathbf{Z},\bm{\delta}}(\mathbf{Z},\bm{\delta}|\widetilde{\bm{\psi}})^{-1}$ with respect to a normal density. Therefore, we have shown \eqref{cond1} for a Bernoulli case. For the Poisson response case, we can represent \eqref{cond1bound3} as 
\begin{equation}
c\exp(2\mathbf{Z}'\mathbf{X}\bm{\beta})\int \Big[\sum_{i=1}^{n}x_{ij}(Z_i - \exp(X_i\bm{\beta} + M_i\bm{\delta})) \Big]^2\frac{\exp(2\mathbf{Z}'\mathbf{M}\delta)f_{\mathbf{Z},\bm{\delta}}(\mathbf{Z},\bm{\delta}|\widetilde{\bm{\psi}})^{-1}}{\exp(\bm{\delta}'\bm{\delta}/\bm{\theta})}
d\bm{\delta}
\label{poibound}
\end{equation}
By the triangle inequality and with some algebra
\begin{equation}
\begin{split}
& c\exp(2\mathbf{Z}'\mathbf{X}\bm{\beta})
\int \Big[\sum_{i=1}^{n}-x_{ij}\exp\Big(-\bm{\delta}'\bm{\delta}/2\bm{\theta} + \mathbf{Z}'\mathbf{M}\bm{\delta} + X_i\bm{\beta} + M_i\bm{\delta} \Big) + \\ 
& x_{ij}z_i\exp\Big(-\bm{\delta}'\bm{\delta}/2\bm{\theta} + \mathbf{Z}'\mathbf{M}\bm{\delta} \Big)\Big]^2f_{\mathbf{Z},\bm{\delta}}(\mathbf{Z},\bm{\delta}|\widetilde{\bm{\psi}})^{-1}d\bm{\delta}  \\
& \leq c\exp(2\mathbf{Z}'\mathbf{X}\bm{\beta})
\int \Big[\sum_{i=1}^{n}|x_{ij}|\exp\Big(-\bm{\delta}'\bm{\delta}/2\bm{\theta} + \mathbf{Z}'\mathbf{M}\bm{\delta} + X_i\bm{\beta} + M_i\bm{\delta} \Big) + \\ 
& |x_{ij}z_i|\exp\Big(-\bm{\delta}'\bm{\delta}/2\bm{\theta} + \mathbf{Z}'\mathbf{M}\bm{\delta} \Big)\Big]^2f_{\mathbf{Z},\bm{\delta}}(\mathbf{Z},\bm{\delta}|\widetilde{\bm{\psi}})^{-1}d\bm{\delta}.
\end{split}
\label{poibound1}
\end{equation}
Similar to a Bernoulli case, \eqref{poibound1} is the expectation of the bounded function $f_{\mathbf{Z},\bm{\delta}}(\mathbf{Z},\bm{\delta}|\widetilde{\bm{\psi}})^{-1}$ with respect to a normal density. Therefore, we have shown \eqref{cond1} for a Poisson case  as well. 

For \eqref{cond2} we can write
\begin{equation}
\begin{split}
\Var\Big(\frac{\partial \log f_{\bm{\delta}}(\bm{\delta}|\bm{\theta})}{\partial \bm{\theta}}  \frac{f_{\mathbf{Z},\bm{\delta}}(\mathbf{Z},\bm{\delta}|\bm{\psi})}{f_{\mathbf{Z},\bm{\delta}}(\mathbf{Z},\bm{\delta}|\widetilde{\bm{\psi}})} \Big) & < 
\E\Big[\Big(\frac{\partial \log f_{\bm{\delta}}(\bm{\delta}|\bm{\theta})}{\partial \bm{\theta}}  \frac{f_{\mathbf{Z},\bm{\delta}}(\mathbf{Z},\bm{\delta}|\bm{\psi})}{f_{\mathbf{Z},\bm{\delta}}(\mathbf{Z},\bm{\delta}|\widetilde{\bm{\psi}})} \Big)^2\Big] \\
& = \int \Big[ \frac{\partial \log f_{\bm{\delta}}(\bm{\delta}|\bm{\theta})}{\partial \bm{\theta}}   \Big]^2
\frac{[f_{\mathbf{Z}|\bm{\delta}}(\mathbf{Z}|\beta_j,\mathbf{M},\bm{\delta})]^2[f_{\bm{\delta}}(\bm{\delta}|\bm{\theta})]^2}{f_{\mathbf{Z},\bm{\delta}}(\mathbf{Z},\bm{\delta}|\widetilde{\bm{\psi}})} 
d\bm{\delta}.
\end{split}
\label{cond2bound1}
\end{equation}
By plugging in the densities for $f_{\bm{\delta}}(\bm{\delta}|\bm{\theta})$ and $f_{\mathbf{Z}|\bm{\delta}}(\mathbf{Z}|\beta_j,\mathbf{M},\bm{\delta})$, we have
\begin{equation}
\int \Big[ \frac{\partial \log f_{\bm{\delta}}(\bm{\delta}|\bm{\theta})}{\partial \bm{\theta}}   \Big]^2
\frac{\exp(2\mathbf{Z}'\eta)f_{\mathbf{Z},\bm{\delta}}(\mathbf{Z},\bm{\delta}|\widetilde{\bm{\psi}})^{-1}}{\exp(2b(\bm{\eta}))\exp(\bm{\delta}'\bm{\delta}/\bm{\theta})}  
d\bm{\delta} = \int \Big[ -\frac{m}{2\bm{\theta}} + \frac{\bm{\delta}'\bm{\delta}}{2\bm{\theta}} \Big]^2
\frac{\exp(2\mathbf{Z}'\eta)f_{\mathbf{Z},\bm{\delta}}(\mathbf{Z},\bm{\delta}|\widetilde{\bm{\psi}})^{-1}}{\exp(2b(\bm{\eta}))\exp(\bm{\delta}'\bm{\delta}/\bm{\theta})}  
d\bm{\delta}.
\label{cond2bound2}
\end{equation}
Since $0 \leq \exp(-2b(\bm{\eta}) )\leq 1$ for both Bernoulli and Poisson cases, \eqref{cond2bound2} is bounded by  
\begin{equation}
\int \Big[ -\frac{m}{2\bm{\theta}} + \frac{\bm{\delta}'\bm{\delta}}{2\bm{\theta}} \Big]^2
\frac{\exp(2\mathbf{Z}'\eta)f_{\mathbf{Z},\bm{\delta}}(\mathbf{Z},\bm{\delta}|\widetilde{\bm{\psi}})^{-1}}{\exp(\bm{\delta}'\bm{\delta}/\bm{\theta})}d\bm{\delta} =
\int \Big[ -\frac{m}{2\bm{\theta}} + \frac{\bm{\delta}'\bm{\delta}}{2\bm{\theta}} \Big]^2
f_{\mathbf{Z},\bm{\delta}}(\mathbf{Z},\bm{\delta}|\widetilde{\bm{\psi}})^{-1}
\exp(-\bm{\delta}'\bm{\delta}/\bm{\theta} + 2\mathbf{Z}'\mathbf{M}\bm{\delta})
d\bm{\delta}.
\label{cond2bound3}
\end{equation}
Therefore, \eqref{cond2bound3} is the expectation of 
\[
\Big[ -\frac{m}{2\bm{\theta}} + \frac{\bm{\delta}'\bm{\delta}}{2\bm{\theta}} \Big]^2
f_{\mathbf{Z},\bm{\delta}}(\mathbf{Z},\bm{\delta}|\widetilde{\bm{\psi}})^{-1}
\]
with respect to a normal distribution. \eqref{cond2}. In conclusion, this proves a finite variance for the numerator part of \eqref{gradientMCL}, which implies a CLT. 

\subsection{A Finite Variance for the Denominator Part}

We need to show 
\begin{equation}
\Var\Big( \frac{f_{\mathbf{Z},\bm{\delta}}(\mathbf{Z},\bm{\delta}^{(k)}|\bm{\psi})}{f_{\mathbf{Z},\bm{\delta}}(\mathbf{Z},\bm{\delta}^{(k)}|\widetilde{\bm{\psi}})} \Big)< \infty
\label{denomcond1}
\end{equation}
Similar to the previous section, we can represent
\begin{equation}
\begin{split}
\Var\Big( \frac{f_{\mathbf{Z},\bm{\delta}}(\mathbf{Z},\bm{\delta}^{(k)}|\bm{\psi})}{f_{\mathbf{Z},\bm{\delta}}(\mathbf{Z},\bm{\delta}^{(k)}|\widetilde{\bm{\psi}})} \Big) & <
\E\Big[\Big( \frac{f_{\mathbf{Z},\bm{\delta}}(\mathbf{Z},\bm{\delta}^{(k)}|\bm{\psi})}{f_{\mathbf{Z},\bm{\delta}}(\mathbf{Z},\bm{\delta}^{(k)}|\widetilde{\bm{\psi}})} \Big)^2\Big] \\
& = \int \frac{[f_{\mathbf{Z}|\bm{\delta}}(\mathbf{Z}|\beta_j,\mathbf{M},\bm{\delta})]^2[f_{\bm{\delta}}(\bm{\delta}|\bm{\theta})]^2}{f_{\mathbf{Z},\bm{\delta}}(\mathbf{Z},\bm{\delta}|\widetilde{\bm{\psi}})} 
d\bm{\delta}.
\label{denomcond1bound1}
\end{split}
\end{equation}
By plugging in the densities for $f_{\bm{\delta}}(\bm{\delta}|\bm{\theta})$ and $f_{\mathbf{Z}|\bm{\delta}}(\mathbf{Z}|\beta_j,\mathbf{M},\bm{\delta})$, we have
\begin{equation}
\begin{split}
c\int \frac{\exp(2\mathbf{Z}'\eta)f_{\mathbf{Z},\bm{\delta}}(\mathbf{Z},\bm{\delta}|\widetilde{\bm{\psi}})^{-1}}{\exp(2b(\bm{\eta}))\exp(\bm{\delta}'\bm{\delta}/\bm{\theta})}
d\bm{\delta} & =
c\exp(2\mathbf{Z}'\mathbf{X}\bm{\beta})\int \frac{\exp(2\mathbf{Z}'\mathbf{M}\bm{\delta})f_{\mathbf{Z},\bm{\delta}}(\mathbf{Z},\bm{\delta}|\widetilde{\bm{\psi}})^{-1}}{\exp(2b(\bm{\eta}))\exp(\bm{\delta}'\bm{\delta}/\bm{\theta})}
d\bm{\delta}\\
& \leq c\exp(2\mathbf{Z}'\mathbf{X}\bm{\beta})\int \frac{\exp(2\mathbf{Z}'\mathbf{M}\bm{\delta})f_{\mathbf{Z},\bm{\delta}}(\mathbf{Z},\bm{\delta}|\widetilde{\bm{\psi}})^{-1}}{\exp(\bm{\delta}'\bm{\delta}/\bm{\theta})}
d\bm{\delta} \\
& = \leq c\exp(2\mathbf{Z}'\mathbf{X}\bm{\beta})\int 
f_{\mathbf{Z},\bm{\delta}}(\mathbf{Z},\bm{\delta}|\widetilde{\bm{\psi}})^{-1}\exp(-\bm{\delta}'\bm{\delta}/\bm{\theta} + 2\mathbf{Z}'\mathbf{M}\bm{\delta})
d\bm{\delta}.
\end{split}
\label{denomcond1bound2}
\end{equation}
This is because the  $0 \leq \exp(-2b(\bm{\eta}) )\leq 1$ in Bernoulli and Poisson cases. Then, \eqref{denomcond1bound2} is the expectation of the bounded function $f_{\mathbf{Z},\bm{\delta}}(\mathbf{Z},\bm{\delta}|\widetilde{\bm{\psi}})^{-1}$ with respect to a normal distribution. Therefore, this proves a finite variance for the denominator part of \eqref{gradientMCL}, which implies a CLT.

\clearpage
\section{Random Projection Methods}
~~~~~We provide an outline of the random projection methods. (Here, we suppress the dependencies of $\mathbf{R}$ on $\phi$.) Let $\mathbf{\Phi}=[\mathbf{I}_{m \times m}, \mathbf{0}_{m \times (n-m)}]'$ be an $n \times m$ truncation matrix. The deterministic {N}ystr{\"o}m's method partitions $n \times n$ positive semi-definite matrix $\mathbf{R}$ into four blocks $\begin{bmatrix}
    \mathbf{R}_{11} & \mathbf{R}_{12} \\
    \mathbf{R}_{21} & \mathbf{R}_{22}
  \end{bmatrix}$, where $\mathbf{R}_{11}=\mathbf{\Phi}'\mathbf{R}\mathbf{\Phi}$. It conducts an exact eigendecomposition of the $m \times m$ sub-matrix $\mathbf{R}_{11}$, where  $\mathbf{V}_{11}$ is its eigenvectors and $\mathbf{\Lambda_{11}}=$ diag$(\lambda_{11,1},\dots,\lambda_{11,m})$ is the diagonal matrix with corresponding eigenvalues. Then, the deterministic {N}ystr{\"o}m's method maps the low-dimensional $m \times m$ eigenvectors $\mathbf{V}_{11}$ to high-dimensional $n \times m$ projection matrix via $\sqrt{\frac{m}{n}}[\mathbf{R\Phi}][\mathbf{V}_{11}\mathbf{\Lambda_{11}}^{-1}]$.

In the probabilistic version, \cite{guan2018computationally} replace the truncation matrix $\mathbf{\Phi}$ with $\mathbf{R}\mathbf{\Omega}$, where $\mathbf{\Omega}$ is an $n \times (m+l)$ random matrix with $\Omega_{ij} \sim N(0,1/\sqrt{m+l})$, where $l$ is an oversampling factor to improve approximation \citep{halko2011finding}. We use $l=m$ as in \cite{guan2018computationally}. To guarantee orthogonality among eigenvectors \cite{guan2018computationally} take a variant of the{N}ystr{\"o}m's method: use the first $m$ columns of the left singular vectors of $[\mathbf{R}\mathbf{\Phi}][\mathbf{V}_{11}\mathbf{\Lambda_{11}}^{-1/2}]$ as the final approximation to the eigenvectors. The algorithm may be simply written as Algorithm~\ref{Projalg}. See \cite{banerjee2012efficient,guan2018computationally} for details. 

\begin{algorithm}
\caption{Random projection method for continuous domain}\label{Projalg}
\begin{algorithmic}[H]
\normalsize
\State Given $n \times n$ positive semi-definite matrix $\mathbf{R}$.

\State 1. Construct and $n\times 2m$ random truncation matrix: 

$\mathbf{\Phi}=\mathbf{R}\mathbf{\Omega}$, where $\Omega_{ij} \sim N(0,1/\sqrt{m+l})$.

\State 2. Conduct the {N}ystr{\"o}m's approximation for eigendecomposition:

Form $\mathbf{R}_{11}=\mathbf{\Phi}'\mathbf{R}\mathbf{\Phi}$

SVD for $\mathbf{R}_{11}=\mathbf{V}_{11}\mathbf{\Lambda_{11}}\mathbf{V}_{11}'$

Form {N}ystr{\"o}m's approximation $\mathbf{C}=[\mathbf{R}\mathbf{\Phi}][\mathbf{V}_{11}\mathbf{\Lambda_{11}}^{-1/2}]$

SVD for $\mathbf{C}=\mathbf{UDV'}$

\State First $m$ columns of $\mathbf{U}$, and first $m$ diagnoal elements of $\mathbf{D}^2$ are used as eigencomponents of the $\mathbf{R}$.

\end{algorithmic}
\end{algorithm}

\section{MCMC for Simulating Random Effects}

~~~~~To simulate random effects in the MCML algorithms, we use the Metropolis-Hastings (MH) whose stationary distribution is $f_{\bm{\delta}|\mathbf{Z}}(\bm{\delta}|\mathbf{Z},\widetilde{\bm{\psi}})$. 
Since projection-based models reduce the dimension of random effects as well as reduce the correlation among random effects, the constructed MCMC algorithm is efficient than that of traditional SGLMMs. We use a standard MCMC algorithm with a multivariate normal proposal function $q(\cdot|\bm{\delta})=MVN(\bm{\delta}, \sigma^2I)$ (e.g. $\sigma^2 = 0.1$). The algorithm may be described as follows: 

\begin{algorithm}
\caption{MCMC for simulating random effects}\label{MCMCrandom}
\begin{algorithmic}[H]
\normalsize
\State  Given a fixed $\widetilde{\bm{\psi}} = ( \widetilde{\bm{\beta}}, \widetilde{\sigma}^2,\widetilde{\phi} )$ and $\bm{\delta}^{(k)}$ at $k$th iteration.

\State 1. Propose  $\bm{\delta}' \sim q(\cdot|\bm{\delta}^{(k)})$.

\State 2. Accept $\bm{\delta}^{(k+1)}=\bm{\delta}'$ with probability

$$\alpha = \min\left\lbrace 1,  \frac{f_{\bm{\delta}|\mathbf{Z}}(\bm{\delta}'|\mathbf{Z},\widetilde{\bm{\psi}})q(\bm{\delta}^{(k)}|\bm{\delta}')}{f_{\bm{\delta}|\mathbf{Z}}(\bm{\delta}^{(k)}|\mathbf{Z},\widetilde{\bm{\psi}})q(\bm{\delta}'|\bm{\delta}^{(k)})} \right\rbrace$$

\State else reject (set $\bm{\delta}^{(k+1)}=\bm{\delta}^{(k)}$).

\end{algorithmic}
\end{algorithm}

\clearpage
\section{Complexity Calculations}

Here we provide details about calculating computational complexity for the algorithms. We summarize our notation in Table~\ref{notation}. All other notations follow previous sections.

\begin{table}
\centering
\begin{tabular}{cc}
  \hline
Symbol & Definition\\
  \hline
$n$ & the number of data points\\
$m$ & reduced rank ($m<<n$) \\
$p$ & the number of covariates \\
$\mathbf{M}$ & $n \times m$ projection matrix\\
$\mathbf{W}$ & $n \times 1$ random effect \\
$\bm{\delta}$ & $m \times 1$ dimension reduced random effect \\
$\mathbf{Z}$ & $n \times 1$ observed non-Gaussian data \\
$\mathbf{X}$ & $n \times p$ design matrix including spatial covariates \\
$\mathbf{Q}$ & $n \times n$ neighbor matrix for discrete case\\
$E[\mathbf{Z}|\bm{\beta}, \mathbf{M}, \bm{\delta}])$ & $n \times 1$ conditional expectation \\
$V(\mathbf{Z}|\bm{\beta}, \mathbf{M}, \bm{\delta})$ & $n \times n$ conditional variance matrix \\
$\mathbf{C}_{\bm{\theta}}$ & $n \times n$ covariance matrix for random effects (continous case) \\
   \hline
\end{tabular}
\caption{Summary of notations used for complexity calculations}
\label{notation} 
\end{table}

\subsection{Continuous Spatial Domain}

\begin{enumerate}
\item Fast MCML approaches

\begin{itemize}
\item Importance sampling step (evaluating $f_{\bm{\delta}|\mathbf{Z}}(\bm{\delta}'|\mathbf{Z},\widetilde{\bm{\psi}})$)\\
matrix multiplications for $\mathbf{M}_{\phi}\bm{\delta}$ and $\bm{\delta}'\bm{\delta}$: $\mathcal{O}(nm) +\mathcal{O}(m)$

\item Update $\bm{\beta},\sigma^2$ (Newton-Raphson)\\
In $\nabla \widehat{l}(\bm{\psi})$ calculations, matrix multiplications for $\mathbf{X}'(\mathbf{Z}-E[\mathbf{Z}|\bm{\beta}, \mathbf{M}, \bm{\delta}])$ and $\bm{\delta}'\bm{\delta}$:\\
$\mathcal{O}(np) + \mathcal{O}(m)$

In $\nabla^2 \widehat{l}(\bm{\psi})$ calculations, matrix multiplications for $-\mathbf{X}'V(\mathbf{Z}|\bm{\beta}, \mathbf{M}, \bm{\delta})\mathbf{X}$: $\mathcal{O}(np(n+p))$

\item Update $\phi$\\
In Algorithm~1, SVD for an $2m \times 2m$ matrix $\mathbf{R}_{11}$: $\mathcal{O}(8m^3)$

In Algorithm~1, SVD for an $n \times 2m$ matrix $\mathbf{C}$: $\mathcal{O}(4nm^2)$

\end{itemize}

\item Standard MCML approaches \citep{christensen2004monte} 

\begin{itemize}
\item Importance sampling step (evaluating $f_{\bm{\delta}|\mathbf{Z}}(\bm{\delta}'|\mathbf{Z},\widetilde{\bm{\psi}})$): $\mathcal{O}(n^2)$\\
matrix multiplications for $\mathbf{W}'\mathbf{C}^{-1}_{\bm{\theta}}\mathbf{W}$

\item Update $\bm{\beta},\sigma^2$ (Newton-Raphson):\\
In $\nabla \widehat{l}(\bm{\psi})$ calculations, matrix multiplications for $\mathbf{X}'(\mathbf{Z}-E[\mathbf{Z}|\bm{\beta}, \mathbf{M}, \bm{\delta}])$ and $\mathbf{W}'\mathbf{C}^{-1}_{\bm{\theta}}\mathbf{W}$: $\mathcal{O}(np) + \mathcal{O}(n^2)$

In $\nabla^2 \widehat{l}(\bm{\psi})$ calculations, matrix multiplications for $-\mathbf{X}'V(\mathbf{Z}|\bm{\beta}, \mathbf{M}, \bm{\delta})\mathbf{X}$: $\mathcal{O}(np(n+p))$

\item Update $\phi$\\
matrix inversion for $\mathbf{C}_{\bm{\theta}}$: $\mathcal{O}(n^3)$

\end{itemize}
\end{enumerate}

\subsection{Discrete Spatial Domain}

\begin{enumerate}
\item Fast MCML approaches

\begin{itemize}
\item Importance sampling step (evaluating $f_{\bm{\delta}|\mathbf{Z}}(\bm{\delta}'|\mathbf{Z},\widetilde{\bm{\psi}})$)\\
matrix multiplications for $\mathbf{M}\bm{\delta}$ and $\bm{\delta}'\mathbf{M}'\mathbf{Q}\mathbf{M}\bm{\delta}$: $\mathcal{O}(nm) + \mathcal{O}(m^2)$

\item Update $\bm{\beta},\tau$ (Newton-Raphson)\\
In $\nabla \widehat{l}(\bm{\psi})$ calculations, matrix multiplications for $\mathbf{X}'(\mathbf{Z}-E[\mathbf{Z}|\bm{\beta}, \mathbf{M}, \bm{\delta}])$ and $\bm{\delta}'\mathbf{M}'\mathbf{Q}\mathbf{M}\bm{\delta}$:\\
$\mathcal{O}(np) + \mathcal{O}(m^2)$

In $\nabla^2 \widehat{l}(\bm{\psi})$ calculations, matrix multiplications for $-\mathbf{X}'V(\mathbf{Z}|\bm{\beta}, \mathbf{M}, \bm{\delta})\mathbf{X}$: $\mathcal{O}(np(n+p))$

\end{itemize}

\item Standard MCML approaches \citep{christensen2004monte} 

\begin{itemize}
\item Importance sampling step (evaluating $f_{\bm{\delta}|\mathbf{Z}}(\bm{\delta}'|\mathbf{Z},\widetilde{\bm{\psi}})$): $\mathcal{O}(n^2)$\\
matrix multiplications for $\mathbf{W}'\mathbf{Q}\mathbf{W}$

\item Update $\bm{\beta},\tau$ (Newton-Raphson)\\
In $\nabla \widehat{l}(\bm{\psi})$ calculations, matrix multiplications for $\mathbf{X}'(\mathbf{Z}-E[\mathbf{Z}|\bm{\beta}, \mathbf{M}, \bm{\delta}])$ and $\mathbf{W}'\mathbf{Q}\mathbf{W}$:\\
$\mathcal{O}(np) + \mathcal{O}(n^2)$

In $\nabla^2 \widehat{l}(\bm{\psi})$ calculations, matrix multiplications for $-\mathbf{X}'V(\mathbf{Z}|\bm{\beta}, \mathbf{M}, \bm{\delta})\mathbf{X}$: $\mathcal{O}(np(n+p))$
\end{itemize}
\end{enumerate}

\bibliography{Reference}
\end{document}